\documentclass[aip,reprint]{revtex4-1}

\usepackage[english]{babel}
\usepackage[%
colorlinks=true,
linkcolor=blue,
citecolor=blue,
urlcolor=blue
]{hyperref}

\usepackage{mathtools}
\usepackage{amsmath}
\usepackage{amsthm}
\usepackage{amsfonts}
\usepackage{float}
\usepackage{graphicx}
\usepackage{caption}
\usepackage{subcaption}
\usepackage{physics}
\usepackage{url}
\usepackage{epstopdf}
\usepackage{lipsum}
\usepackage[title]{appendix}
\apptocmd{\appendices}{\apptocmd{\thesection}{: }{}{}}{}{}
\usepackage[capitalise,sort&compress]{cleveref}
\usepackage{titlesec}
\usepackage{gensymb}
\usepackage[separate-uncertainty = true]{siunitx}
\usepackage{colortbl}
\usepackage{wrapfig}
\usepackage{setspace}
\usepackage{amssymb}
\usepackage[ampersand]{easylist}
\usepackage{array}
\usepackage{cancel}
\usepackage[T1]{fontenc}
\usepackage[utf8]{inputenc}
\usepackage[flushleft]{threeparttable}
\usepackage{soul}
\usepackage{xcolor}
\sethlcolor{yellow}
\usepackage[margin=0.6in]{geometry}
\usepackage{lmodern}
\crefname{equation}{Eq.}{Eqs.}
\crefname{figure}{Fig.}{Figs.}
\crefname{tabular}{Table}{Tables}
\crefname{algocf}{Algorithm}{Algorithms}
\crefname{remark}{Remark}{Remarks}
\crefname{theorem}{Theorem}{Theorems}

\Crefname{equation}{Equation}{Equations}
\Crefname{figure}{Figure}{Figures}
\Crefname{tabular}{Table}{Tables}
\Crefname{algocf}{Algorithm}{Algorithms}
\Crefname{remark}{Remark}{Remarks}
\Crefname{theorem}{Theorem}{Theorems}

\renewcommand{\arraystretch}{1.25}

\graphicspath{ {./Images/}}

\makeatletter
\def\@email#1#2{%
 \endgroup
 \patchcmd{\titleblock@produce}
  {\frontmatter@RRAPformat}
  {\frontmatter@RRAPformat{\produce@RRAP{*#1\href{mailto:#2}{#2}}}\frontmatter@RRAPformat}
  {}{}
}%
\makeatother

\begin{document}


\title{Full-Field Quantitative Visualization of Shock-Driven Pore Collapse and\\ Failure Modes in PMMA} 



\author{Barry P Lawlor*}
\email{blawlor@caltech.edu}
\affiliation{Division of Engineering and Applied Science, California Institute of Technology, Pasadena, CA 91125, USA}

\author{Vatsa Gandhi}
\affiliation{Department of Engineering, University of Cambridge, Cambridge, CB2 1PZ, UK}

\author{Guruswami Ravichandran}
\affiliation{Division of Engineering and Applied Science, California Institute of Technology, Pasadena, CA 91125, USA}

\date{\today}

\begin{abstract}
    The dynamic collapse of pores under shock loading is thought to be directly related to hot spot generation and material failure, which is critical to the performance of porous energetic and structural materials. However, the shock compression response of porous materials at the local, individual pore scale is not well understood. This study examines, quantitatively, the collapse phenomenon of a single spherical void in PMMA at shock stresses ranging from $0.4-1.0$ GPa. Using a newly developed internal digital image correlation technique in conjunction with plate impact experiments, full-field quantitative deformation measurements are conducted in the material surrounding the collapsing pore for the first time. The experimental results reveal two failure mode transitions as shock stress is increased: (i) the first \textit{in-situ} evidence of shear localization via adiabatic shear banding and (ii) dynamic fracture initiation at the pore surface. Numerical simulations using thermo-viscoplastic dynamic finite element analysis provide insights into the formation of adiabatic shear bands (ASBs) and stresses at which failure mode transitions occur. Further numerical and theoretical modeling indicates the dynamic fracture to occur along the weakened material inside an adiabatic shear band. Finally, analysis of the evolution of pore asymmetry and models for ASB spacing elucidate the mechanisms for the shear band initiation sites, and elastostatic theory explains the experimentally observed ASB and fracture paths based on the directions of maximum shear.
\end{abstract}


\maketitle 

\section{Introduction} \label{sec:Introduction}

Porous materials such as energetic (e.g., polymer bonded explosives) and structural (e.g., foams) materials feature in many applications. When such materials are dynamically loaded (e.g., shocked), a pore collapse phenomenon occurs, the physics of which has great implications for both these material classes. Generally, in energetic materials, the main concern is with mechanically-induced hot spot generation via stress waves, which likely dictates the detonation response of the material \cite{Tarver1996Critical,Zhou2013Ignition,Heavens1974ThinLayer,Coffey1981LocalHeating,Armstrong2012Hotspot}. This has often been linked to pore collapse \cite{Field1982Hotspot,Field1992HotSpot,Bourne2003Temperature}. Thus, it is desirable to understand the mechanical response of pore collapse, which may provide insights to the hot spot generation mechanisms and aid in preventing accidental detonations. 

For structural materials, porosity may be either intentional, as is the case for metallic foams and architected metamaterials, or unintentional through defects in manufacturing processes such as in additive manufacturing. Additionally, in these examples the length-scale of porosity ranges from nanometers through centimeters, making the material response of porous media a rich topic of study. The primary issue with these structures lies in the localized material and structural failure which is often driven by stress concentrations in the neighborhood of pores. For the most effective prediction of failure in these structures, it is critical to characterize the failure response of the material, and to understand it from a fundamental, local perspective at the individual pore scale. Alternatively, some porous media can be treated as continuum materials through the investigation of the effective or homogenized response when the pore scale is far smaller than the application scale.

The macroscopic, or continuum, response of porous materials under dynamic loading has been studied at great length. In general, porous materials have been shown to possess favorable qualities such as shock disruption via micro-inertial effects and energy absorption due to the large plastic work required to fully close pores. However, they generally suffer from lower spall strength, making design with porous materials a complicated task. Theoretical approaches, assuming symmetric collapse of pores, originated with the work of Hermmann which enabled the introduction of porosity to the equation of state for materials in what is known as the $P-\alpha$ theory \cite{Herrmann1969Constitutive}. Carroll and Holt subsequently developed an analytical form for the $\alpha$ parameter \cite{Carroll1972Relation} through the analysis of a thick spherical shell. Following this work, many modifications were made to the $P-\alpha$ theory. For example, Butcher, et al. incorporated the influence of deviatoric stress, work-hardening, initial void size, and material viscosity \cite{Butcher1974Compaction} which revealed the effect of micro-inertia and viscosity on delaying the void collapse. Furthermore, extensive analytical, numerical, and experimental investigations have followed to understand the macroscopic response of porous solids under dynamic loading conditions \cite{Ravi1993PoreCollapseVisco,Czarnota2006Void,Czarnota2017PorousStructure,Lovinger2021Micro}.

While the macroscopic (continuum) response to shock loading, measured through traditional interferometric techniques, has been well investigated to characterize the equation of state and shock structure, studies on the deformation and failure at the local length scale of individual pores have been sparse. However, the advances in high speed imaging technologies in the last few decades have begun to enable full-field investigation at the local scale. Early experiments were limited to qualitative characterization of the collapse of cylindrical holes in transparent gels \cite{Bourne1992Collapse,Dear1988Array} through high speed imaging, and later were extended to quantitative characterization via particle image velocimetry \cite{Swantek2010Arrays}. These authors observed jetting at higher stresses and also clear evidence of shock shielding and amplification effects when various arrays of holes were loaded. 

Recently, access to advanced x-ray synchrotron sources has enabled internal imaging of opaque structural materials during plate impact/shock experiments. Investigation, using x-ray phase contrast imaging (PCI), of additively manufactured lattice structures under shock revealed insights into the jet initiation (simple cubic) and mitigation (face centered tetragonal) phenomena which were controlled by the details of the lattice structures \cite{Branch2017ControllingShock}. Further, the role of material parameters, lattice length scales, and impact velocity on jetting transitions was systematically investigated via experiment and simulation \cite{Lind2020Jetting}. Similar experiments demonstrated that a pore in the shape of an elongated triangle behaved as a shock diode: promoting jetting when impacted from one direction, while disrupting the shock from the other direction \cite{Branch2020Diode}. These experiments also investigated the shock shielding properties of a series of holes. 

A close experimental analog to the spherical pore collapse phenomena is the hole closure experiment, in which specimens with cylindrical holes are dynamically loaded via plate impact. Glazkov, et al. used these experiments, with pre- and post-mortem hole size measurements, to infer material response at high strain rates \cite{Glazkov2009Peculiarities}. Lind, et al. performed similar hole closure experiments in copper \cite{Lind2021Closure}, now using PCI to capture the evolution of the area of the hole, \textit{in-situ} during closure. They primarily used the rich and complicated loading state in these experiments to calibrate material models at high strain rates and large strains. However, the inverse analysis also offers insights into the physics of pore collapse, as the rate of closure and final collapsed volume were shown to be highly dependent on the strain rate-hardening of the material. Further analysis of these experiments could inform parameters for analytical continuum models such as the $P-\alpha$. Follow-up experiments on tantalum \cite{Nelms2022ClosureTantalum} served a similar role for model parameter calibration, while also observing possible shear localization (an important failure and hot-spot generation mechanism) near the closed hole through post-mortem electron backscatter diffraction (EBSD) analysis. 

To investigate the physics of pore collapse, Escauriza, et al. conducted plate impact experiments on PMMA with a spherical pore \cite{Escauriza2020Collapse}. Using PCI they captured the pore collapse evolution---observing a transition from strength dominated (low pressure) to hydrodynamic (high pressure) regimes, the development of cracks at low pressures, and jetting instabilities at high pressures. Complementary numerical simulations predicted the development of adiabatic shear bands during these collapse events \cite{Rai2020PoreCollapse}. Recently, Lovinger and Kositski carried out cylindrical pore collapse experiments in Ti-6Al-4V specimens, leveraging plate impact with a soft-catch setup, which enabled post-mortem analysis of the specimens \cite{Lovinger2024Localization}. Their work revealed the first definitive evidence of shear localization in pore collapse via adiabatic shear banding and shear cracking. They further explored the role of pore size and spacing on failure initiation and connectivity, respectively. While the aforementioned works have all been limited by sample fabrication and experimental resolution to pores near the mm scale, a new technique has been recently developed which uses PCI to capture the collapse of spherical pores on the order of several $10$s of microns \cite{Hodge2021MicroVoid}. This technique promises future understanding of pore collapse at very fine spatial scales which are relevant to many commonly occurring pores. These experimental efforts and complementary computational endeavors have enriched our understanding of the pore collapse phenomena. However, to understand the details of shear localization and failure thresholds, a need still remains for quantitative full-field measurements during the pore collapse phenomena under shock compression.

To quantitatively characterize the deformation field around a pore under shock compression, it is proposed to perform plate impact experiments on PMMA samples with a single, embedded, spherical pore, in conjunction with high speed digital image correlation (DIC) \cite{Sutton2009}---a non-contact quantitative imaging technique. Plate impact experiments require measurements to be performed at the center of the target plate, such that the uniaxial strain condition is maintained throughout the measurement, until the time at which release waves arrive. Hence, the embedded pore must be located at the center of the sample, and DIC performed internally at the mid-plane of the pore. This is essential to capture the true deformation response and failure mechanisms during the pore collapse event, as opposed to alternative approaches of making surface measurements and leveraging computational tools to infer the internal response. Further, this approach enables a more physically relevant investigation of spherical pores undergoing collapse, as opposed to cylindrical holes.

In this study, the internal DIC technique \cite{Lawlor2024IntDIC} for shock compression experiments is implemented to study pore collapse at shock stresses ranging from $0.4-1.0\,$GPa. The internal DIC method is briefly summarized, and details regarding sample preparation, plate impact experiments, and high speed imaging are described in \cref{sec:Methods}. Results for the experiments and initial DIC analysis are presented in \cref{sec:Results}. Simulations are then conducted, replicating the loading conditions of the experiments in \cref{sec:Modeling}, and the physics governing failure mechanisms are discussed and supported with theoretical models in \cref{sec:Discussion}. Lastly, concluding remarks are given in \cref{sec:Conclusion}.

\section{Materials and Methods} \label{sec:Methods}
    \subsection{Quantitative Imaging of Displacements}

	Digital image correlation (DIC) is a non-contact, full-field, quantitative imaging technique, which measures the displacement field of a body undergoing deformation by correlating, or pattern matching, a unique grayscale speckle pattern which is applied to the specimen prior to deformation \cite{Sutton2009}. Images taken before and throughout the deformation then contain the necessary information for correlation to extract the displacement field with respect to the reference (undeformed) configuration, which can be converted to full-field strain measurements through discrete differentiation. The technique is traditionally restricted to surface measurements; however, in this work an internal DIC framework is employed to capture deformation around a collapsing pore under shock compression. The concept of internal DIC has been successfully implemented previously in the quasi-static regime \cite{Berfield2007Micro} and in dynamic laser-induced cavitation experiments in gels, conducted under a microscope \cite{McGhee2023Microcavitation}. Recently, this technique was extended to be used in full-scale dynamic experiments, demonstrating its feasibility and accuracy for application to both split-Hopkinson (Kolsky) pressure bar and plate impact shock experiments. Details of the development of the internal DIC technique are presented elsewhere \cite{Lawlor2024IntDIC}, and are only briefly summarized here. The optical technique involves embedding a speckle pattern at the internal plane of the transparent sample. By visualizing through the transparent sample (shifting the camera's focal plane to the internal plane), in-plane deformation is captured in the images (i.e., the direction of compression is in the same plane as the speckle pattern, in this case, horizontally). Here, normal plate impact experiments were conducted on PMMA samples, coupled with ultra-high speed imaging and digital image correlation (DIC) \cite{Sutton2009,Ravindran2023ThreeDim,Ravindran2023Mesoscale} to extract full-field quantitative strain measurements surrounding the pore during deformation.

    \subsection{Sample Preparation}
        \begin{figure*}[htpb]
            \centering
            \includegraphics[width=1.0\textwidth]{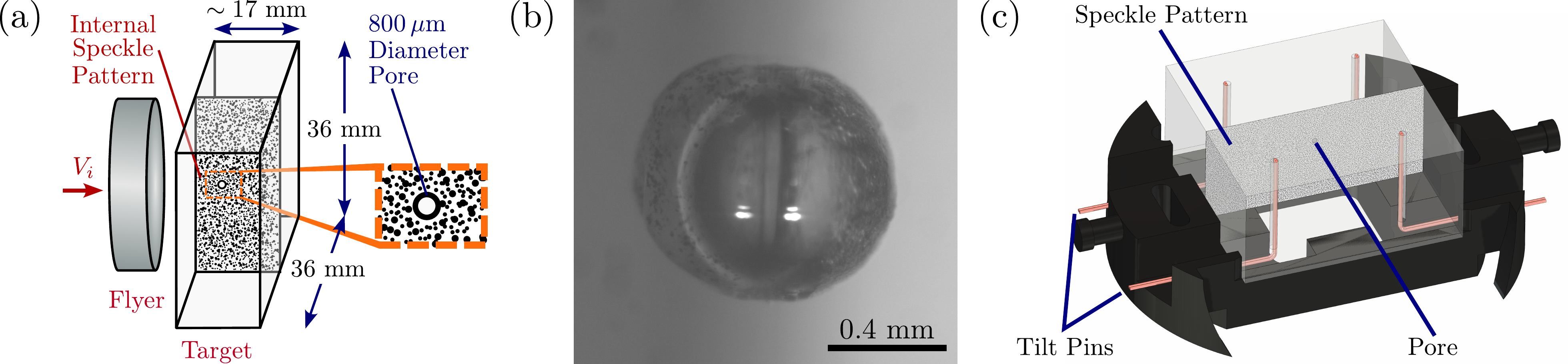}
            \caption{Details of the specimen preparation. (a) Plate impact pore collapse flyer and target specimens. (b) Microscope image of the spherical pore inside the target plate. Glue interface is slightly visible at the center of the pore. (c) CAD schematic of target plate, mounted onto the front of the target holder and equipped with electrical shorting pins. The top surface is the impact plane.}
            \label{fig:specimengeometries}
        \end{figure*}
        
        Specimen preparation for plate impact experiments is a stringent process, requiring precise, detailed steps to ensure the specimens are suitable for the experiments. The target specimens for all experiments in this study, depicted in \cref{fig:specimengeometries}\hyperref[fig:specimengeometries]{a}, were manufactured from PMMA stock material obtained from E\&T Plastics (Long Island City, New York). To embed the internal DIC speckle pattern and spherical pore, the target is manufactured from two cuboidal half-samples with rectangular cross section of nominal dimensions, $36\,$mm $\times$ $18\,$mm and $17\,$mm thick. Both half-samples are carefully lapped together in the impact and visualization directions to ensure identical dimensions and create ideal bonding surfaces. The visualization surfaces (all surfaces which are parallel to the speckle pattern plane) are then mirror polished to maximize transparency, and matching hemispheres of radius $0.4\,$mm are milled into each half-sample such that they precisely align to their counterpart to generate a nearly perfect spherical pore when combined. Next, the DIC speckle pattern is applied to the inner surface of one half-sample by airbrushing a thin layer of transparent paint and blowing toner powder onto the wet paint, before adding a second layer of transparent paint to seal the speckle pattern.

        After the paint is allowed to dry for at least 24 hours, the half-samples are glued together with a two-part epoxy, EpoxAcast 690 (SmoothOn, Macungie, Pennsylvania). Confining plates are used to precisely align the two hemispheres to generate a pristine spherical pore, and weights are applied to the sandwiched sample to squeeze out excess glue and thus maximize the glue bond strength. Once the sample has settled for approximately ten minutes, the weights are removed, the sample is carefully disassembled, and any glue which has filled the pore is removed. The assembly is then reset with confinement and weights, without additional glue, to ensure the pore is filled with air, rather than glue. Finally, after the glue cures for 24 hours, the now-intact sample is removed, re-lapped in the impact direction to meet specifications for the experiment, and the alignment of the pore is inspected under a microscope. \Cref{fig:specimengeometries}\hyperref[fig:specimengeometries]{b} shows an example of the pore alignment. The final restrictions for the target sample are less that $30\,\mu$m variation in thickness and less than $1\,\mu$m surface flatness as measured by Fizeau rings under monochromatic light with a quartz optical flat. After the target is assembled, the final specimens are nominally $36\times36\,$mm square plates which are $17\,$mm thick.

        Flyer plates made of PMMA or aluminum ($7075$ alloy) are more straightforward to manufacture, starting with machined cylindrical samples of $35\,$mm diameter and $13\,$mm thickness. They are lapped until the thickness variation is less than $10\,\mu$m and the surface flatness is less than $0.5\,\mu$m. These restrictions on the target and flyer plates ensure planar shock wave generation upon impact. Finally, the flyer plate is glued into the projectile and inserted in the gun barrel.

        Electrical shorting pins are used in these experiments to trigger diagnostics and to measure the angle of inclination (tilt) between the flyer and target at impact. These pins are inserted into holes in the target plate which correspond with the perimeter of the flyer, and are glued in place, as shown in \cref{fig:specimengeometries}\hyperref[fig:specimengeometries]{c}. They are then sanded down, and lapped until flush with the impact surface. Finally, the target plate is glued to the target holder, the shorting pins are wired into a logic circuit box that is connected to a digital oscilloscope, and the target holder is mounted onto a six-degree-of-freedom gimbal which is moved into the vacuum chamber for alignment.

    \subsection{Plate Impact Experimental Setup}

\begin{figure*}[htpb]
            \centering
            \includegraphics[width=0.8\textwidth]{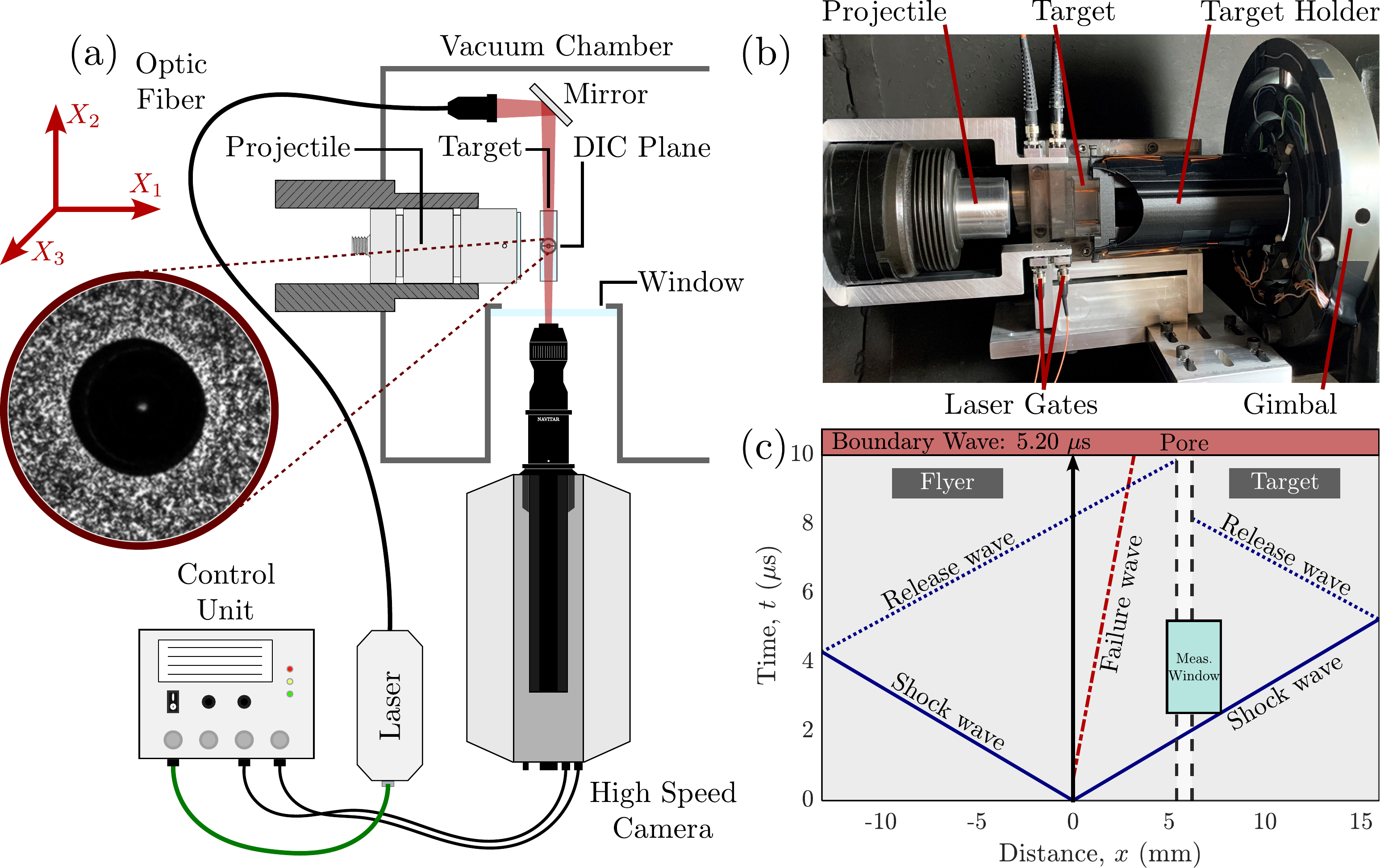}
            \caption{Experimental setup for plate impact experiments. (a) Schematic of experimental components setup in and around the vacuum chamber. Depicted is the projectile with flyer plate affixed, approaching the target plate. Also shown is the high speed, high magnification imaging setup, complete with backlit laser illumination. An example image of the pore and the surrounding region is shown in the inset. (b) Photo of the chamber setup for a shot. Six-DOF gimbal shown on the right side of the image, with the target mounted onto it via a long target holder. Also included is the housing for the laser illumination source behind the target. (c) Distance-time $(x-t)$ diagram for Pore--0.4, used for experimental design to maximize measurement time window. Shock, release, and failure waves are labeled accordingly; the impact plane between flyer and target lies at $x=0$; the pore is marked by two vertical, dashed lines. The ``measurement window'' is visualized by a light blue box, which is bounded in the $x$ dimension to indicate the camera field of view and bounded in the $t$ dimension to indicate the beginning and end of the measurement window. }
            \label{fig:PlateImpactSetup}
        \end{figure*}

        Normal plate impact experiments were performed using a powder gun facility at Caltech, which consists of a $38.7\,$mm diameter keyed barrel of $3\,$m length. Before running the experiment, the projectile is placed at the end of the gun barrel, and brought close to the target, which is mounted to the gimbal in the vacuum chamber. Careful translational and rotational alignment, using an auto-collimator, is performed to ensure perfect alignment between the impact surfaces of the two plates. This is critical to minimize the impact tilt between the flyer and target plates during the shot and ensure plane wave propagation. The high speed camera is also optimally configured and aligned parallel to the target visualization surface closest to the camera, after which the field of view and focal plane are set. A schematic of the full experimental setup is provided in \cref{fig:PlateImpactSetup}\hyperref[fig:PlateImpactSetup]{a}, and an image of the setup inside the vacuum chamber is shown in \cref{fig:PlateImpactSetup}\hyperref[fig:PlateImpactSetup]{b}. The experiment begins when the gunpowder charge is detonated behind the projectile, propelling it down the barrel where it impacts the target plate, situated in the vacuum chamber at the end of the barrel. Upon impact, shorting pins trigger the high speed imaging diagnostics to begin recording. Simultaneously, a planar shock wave is generated in both the target and flyer, and the camera captures images of the shock propagation and the in-plane deformation of the target plate. To measure the impact velocity, a laser system composed of two precisely spaced laser gates captures the moment at which the projectile breaks each laser plane. In some experiments, which do not use a conductive flyer (Pore--0.4), the laser gate system is also used to trigger the camera to record.

        Plate impact experiments require careful design to ensure the desired loading state is applied and the material response is measured prior to the arrival of release waves. Flyer plate materials and impact velocities are selected to generate the desired shock stresses in the target material. For a given impact velocity ($V_i$), the imposed particle velocity ($u_p)$ in the target material is determined through the impedance matching technique \cite{Meyers1994DBOM} based on the known ambient material density ($\rho_0$) and equation of state ($U_s-u_p$ relation, where $U_s$ is the shock wave speed, given in \cref{eq:EOS}) for both the flyer and target materials. From there, shock stress is calculated using \cref{eq:Impedance}. \begin{equation} \sigma_{11} = \rho_0 U_s u_p \label{eq:Impedance} \end{equation} 

The timing of several release waves must also be calculated, including those initiating from the flyer free surface, the target free surface, and the perimeter/boundary of the target plate because of traction-free conditions. The design can be visualized through a distance-time $(x-t)$ diagram which plots the propagation of all relevant stress waves in the experiment. An example $x-t$ diagram is shown in \cref{fig:PlateImpactSetup}\hyperref[fig:PlateImpactSetup]{c} for the $0.41\,$GPa shock stress pore collapse experiment. Upon impact at $\text{time, }t=0$, a normal shock wave is generated in both the target (forward-propagating) and flyer (backward-propagating), which induces pore collapse after passing the pore at $2.03\,\mu$s. The ``measurement window'' begins when the shock wave passes the camera field of view at $2.54\,\mu$s and ends when the field of view is disturbed by either a free surface release, boundary release, or the arrival of the ``failure wave'' observed previously in PMMA \cite{Huang2015FailureWave}. These waves alter the uniaxial strain loading conditions, occlude the field of view, or both. This measurement window can be maximized for a given shock stress by using a thick flyer and target (to delay the arrival of free surface release waves), and optimizing the distance of the pore away from the impact surface to balance competition between boundary wave and failure wave arrival. Flyer diameter and target width could also be used to delay boundary wave arrival, but are limited by the size of the powder gun barrel. In the example used in \cref{fig:PlateImpactSetup}\hyperref[fig:PlateImpactSetup]{c}, the measurement window is closed by the arrival of boundary release waves at $5.20\,\mu$s producing a predicted window of $2.66\,\mu$s for Pore--0.4. Unfortunately, the failure wave speed increases with shock stress, making experiment design difficult at higher stresses by significantly shortening the maximum measurement window.

    \subsection{High Speed Imaging and Digital Image Correlation (DIC)} \label{sec:DIC}

        To capture clear images of the speckle pattern during the dynamic event, precise visualization setup is necessary. The high speed imaging setup is composed of the HPV-X2 camera (Shimadzu, Kyoto, Japan) set to capture images at $10,000,000$ frames/second ($100\,$ns inter-frame time, $50\,$ns exposure time) along with the CAVILUX Smart pulsed, incoherent laser light source (Cavitar, Tampere, Finland) which is synchronized to the camera exposures and pulsed for $50\,$ns durations. To perform high magnification imaging with sufficient resolution to capture fine features of deformation at the internal plane, the camera is equipped with a $0.7-4.5$x zoom lens and 2x adapter tube (Navitar, Rochester, New York), which achieved a typical field of view of $2.76\times1.73\,$mm ($400\times250\,$pixels). For such a small field of view, the camera is also mounted to a five-degree-of-freedom optical stage which allows precise alignment of the camera's field of view. The high magnification lens also produces a highly light-starved environment, which is remedied by configuring the laser light source in a back-lit setup to maximize illumination. Additionally, the large curvature of the zoom optic generates image distortions, which are addressed through standard distortion correction procedures. Images are taken of the rigid body translation of the target at several vertical and horizontal distances, and a correction function is computed which regains uniform displacement fields. This correction function is subsequently used for the experimental images as well. Finally, as was discussed in previous work \cite{Lawlor2024IntDIC}, some optical distortions arise resulting from the change in refractive index across the shock front. This is mitigated by aligning the camera lens to the target visualization surface closest to the camera to maximize the parallelism between the two.

        Having established a basis for capturing clear images of the speckle pattern in this setting, one must also consider the details of the quantitative imaging technique, DIC. As mentioned above, the internal speckle pattern is generated by randomly distributing toner powder inside a layer of transparent paint, which produces a typical speckle size of $10-20\,\mu$m. This is an ideal size compared to the camera's spatial resolution of $7\,\mu$m/pixel. Correlation of the speckle pattern images is performed with the Vic-2D software \cite{VIC2D} (Correlated Solutions, Columbia, South Carolina) along with the built-in distortion correction algorithm mentioned earlier. For the sake of consistency between experiments, the same correlation parameters are used to post-process each experiment: 21 pixel subset size (SS), 1 pixel step size (ST), and strain is computed using a 15 pixel, 90\% center-weighted, Gaussian strain window (SW). Further, the effects of the subset size, step size, and strain window size can be summarized through the virtual strain gage length \cite{Reu2015VSG} $\left(L_{\text{VSG}}\right)$, computed as \begin{equation} L_{\text{VSG}} = (SW-1)ST + SS \label{eq:VSG} \end{equation} which is representative of the total filtering applied to the physical features during the DIC analysis. This parameter is used in \cref{sec:Discussion,sec:AppendixDIC} to discuss DIC measurements of fine features such as shear bands.

\section{Results} \label{sec:Results}
    Four pore collapse experiments were conducted at shock stresses ranging from $0.4-1.0\,$GPa. The shock conditions were controlled primarily by varying the impact velocity, as well as changing the flyer plate material as-needed. For all the experiments, the pore size was kept constant at $800\,\mu$m diameter while the shock stress was varied. Details of the specimen geometries and loading conditions are summarized in \cref{tab:Plate Impact Shot Summary}. The naming convention adopted for numbering the experiments indicates the experiment type and nominal shock stress (e.g., Pore--0.6 refers to the pore collapse experiment under nominally $0.6\,$GPa shock stress).

    \begin{table*}[ht]
    \setlength{\tabcolsep}{7.5pt}
    \centering
    \caption{\label{tab:Plate Impact Shot Summary} Summary of pore collapse experiments.}
    \resizebox{1.0\textwidth}{!}{
    \begin{tabular}{c c c c c c c c} \hline\hline
     Shot                   & Flyer         &       Flyer               & \phantom{**}Target**               & \phantom{***}L$_{\text{Pore}}$***    & Impact Velocity   & Shock Stress          & Tilt\\
     Number                 & Material      &       Thickness (mm)      & Thickness (mm)       & (mm)                               & $V_i$ (m/s)       & $\sigma_{11}$ (GPa)   & (mrad)\\
    \hline
    Pore--0.4               & PMMA          & $13.069 \pm 0.002$        & $15.938 \pm 0.027$   & 5.4                                & $227 \pm 7$      & $0.41 \pm 0.01$       & N/A\\
    \phantom{*}Pore--0.6*   & Al 7075       & $12.893 \pm 0.003$        & $22.878 \pm 0.004$   & 7.1                                & $213 \pm 5$      & $0.62 \pm 0.02$       & 0.78\\
    Pore--0.8               & Al 7075       & $12.845 \pm 0.002$        & $15.793 \pm 0.022$   & 5.2                                & $263 \pm 10$     & $0.78 \pm 0.03$       & 3.8\\
    Pore--1.0               & Al 7075       & $12.945 \pm 0.002$        & $16.492 \pm 0.008$   & 5.1                                & $334 \pm 8$      & $0.99 \pm 0.02$       & 3.4\\
    \hline\hline
    \end{tabular}}
        \begin{footnotesize}
        *An Al 7075 buffer plate (thickness $0.601 \pm 0.001\,$mm) was glued to the target impact surface for this experiment.\\
        **All targets plates were manufactured from PMMA.\\
        ***L$_{\text{Pore}}$ indicates the distance from impact surface to the front edge of the pore.
        \end{footnotesize}
    \end{table*}

    \subsection{Deformation Images, Longitudinal and Lateral Strain, and Pore Volume} \label{sec:Results-Deformation}
\begin{figure*}[htpb]
            \centering
            \includegraphics[width=0.95\textwidth]{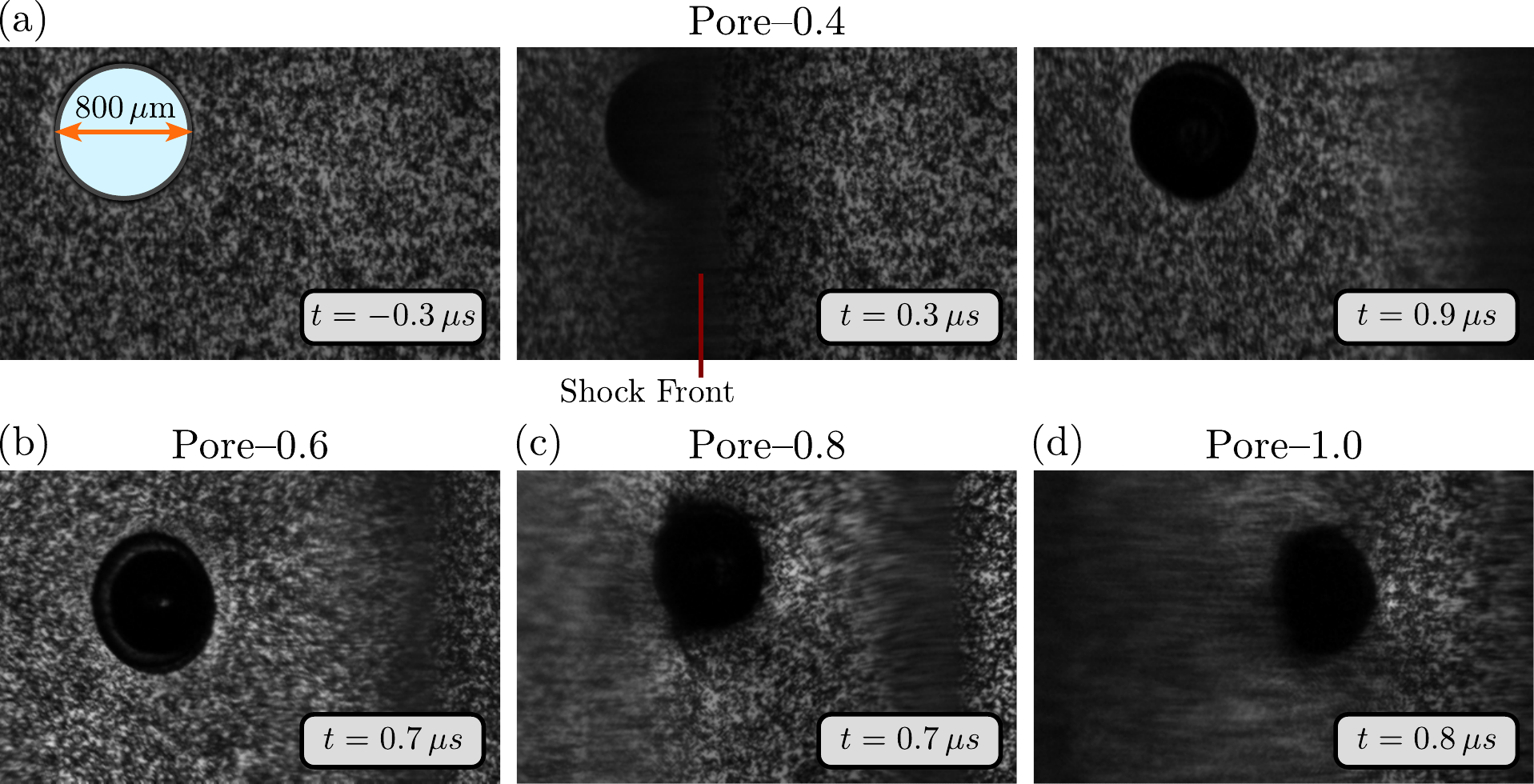}
            \caption{Deformation images at the internal mid-plane for all pore collapse experiments. Each experiment is performed with a single spherical pore that is $800\,\mu$m in diameter. (a) Experiment Pore--0.4. Three time instances are shown: before shock (left), during shock (middle), and after shock (right). (b-d) Pore--0.6, Pore--0.8, and Pore--1.0 after shock loading. $t=0$ corresponds to the time of arrival of the shock at the pore.}
            \label{fig:DeformationImages}
        \end{figure*}

        For each experiment, the entire shock loading process is captured via high speed imaging, beginning with the ambient unshocked state and ending with the arrival of release waves which concludes the measurement window. A representative set of images for all four experiments is shown in \cref{fig:DeformationImages}, and \cref{fig:DeformationImages}\hyperref[fig:DeformationImages]{a} illustrates the evolution of Pore--0.4. From left to right, one can see the unshocked (ambient) condition, the shock wave (dark band) passing the pore (propagating from left to right), and the shocked state. Though not shown here, it is also possible to briefly visualize the radial wave reflection off of the pore surface. Similarly, \cref{fig:DeformationImages}\hyperref[fig:DeformationImages]{b-d} show one frame of the shocked state for each of the remaining experiments, as the details of evolution are generally difficult to ascertain from the deformation images alone. DIC analysis will reveal more insight to the evolution of the shocked state. In general, the deformation images reveal a clear progression from nearly indiscernible collapse in Pore--0.4 to significant collapse in Pore--1.0. They also show development of fracture at the pore interface in Pore--0.8, which will be discussed further in \cref{sec:Results-Fracture}.

\begin{figure*}[htpb]
            \centering
            \includegraphics[width=1.0\textwidth]{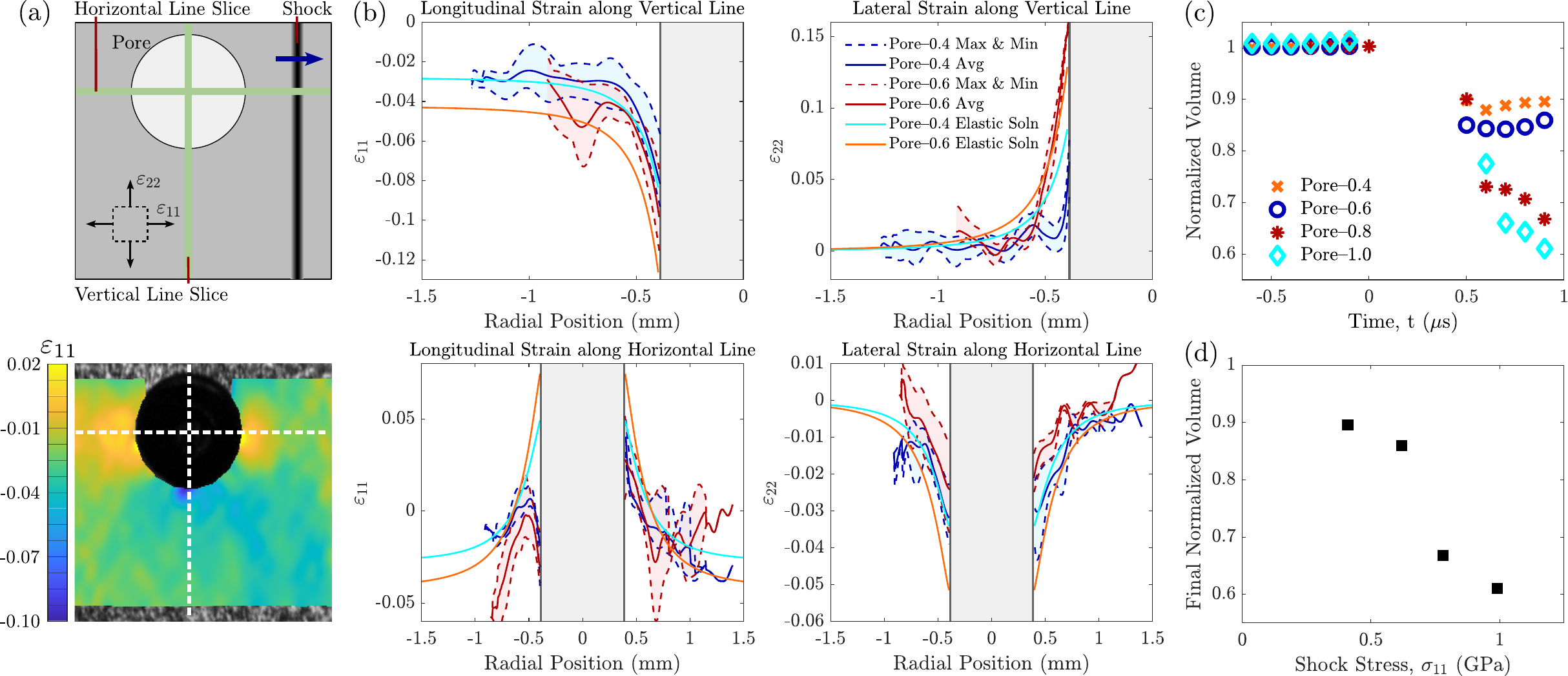}
            \caption{Line slice comparison to elastic solution and pore volume analysis. (a) Schematic of pore, shock direction, and line slices; DIC snapshot of longitudinal strain $\left(\varepsilon_{11}\right)$ with line slice locations overlaid. (b) Experimental strain measurements $\left(\varepsilon_{11}, \, \varepsilon_{22}\right)$ taken along line slices and plotted along with corresponding static elastic solution (\cref{sec:AppendixElastic}). Experimental data is averaged (solid curve) over all time instances after reaching the shocked state and bounded by the minimum and maximum measurements (dashed curves).  (c) Normalized pore volume evolution for each pore collapse experiment. (d) Final measured normalized volume for each experiment.}
            \label{fig:LineSlice}
        \end{figure*}

        After capturing deformation images with a clear speckle pattern after shock arrival, the images were processed via DIC, computing the displacement field and subsequently the Lagrangian strain measures.
        While blurry regions (and fracture, in the case of Pore--0.8) obscured the speckle pattern in Pore--0.8 and Pore--1.0, the image quality in Pore--0.4 and Pore--0.6 enabled excellent correlation. To analyze the strain concentrations around the collapsing pore, line slices are taken vertically and horizontally through the center of the pore. A schematic of this concept is shown in \cref{fig:LineSlice}\hyperref[fig:LineSlice]{a} along with an example using full-field DIC data from Pore--0.4. The results for longitudinal ($\varepsilon_{11}$) and lateral ($\varepsilon_{22}$) strain along both the vertical and horizontal lines are presented in \cref{fig:LineSlice}\hyperref[fig:LineSlice]{b}. Each plot includes the experimental results combined from all time instances after the shocked state was achieved, along with the static, elastic solution for uniaxial strain compression of a spherical pore. This is determined using the elastic solution for static uniaxial stress loading imposed on a spherical pore, first derived by Southwell \cite{Southwell1926Concentration}. The solution is superposed to incorporate the shock stress and lateral confining stress (\cref{sec:AppendixElastic}, \cref{eq:LatStress}) associated with the plate impact experiment's uniaxial strain loading condition. Elastic parameters for PMMA under dynamic compression with confinement are used \cite{Rittel2008DynamicFlow}; further details can be found in \cref{sec:AppendixElastic}.

        The line slice plots reveal good overall agreement between experimental and theoretical results, demonstrating the ability to accurately capture strain concentrations as large as $16\%$. In particular, Pore--0.4, which is nearly completely elastic, coincides very closely with the elastic solution. The quality of the DIC data is particularly good along the vertical line, while more noise is present along the horizontal line. This likely arises due to diffraction and the shock reflections off the pore at the front and back, leading to a complex deformation state. While agreement with the elastic solution provides confidence in the experimental result, deviations from the elastic solution are more instructive. In particular, the longitudinal strain along the horizontal line reveals a significant discrepancy from the elastic solution. The strain at the back of the pore (positive radial position) coincides with the theory, but the front of the pore undergoes much more significant compressive strain in both experiments. This is a clear indication of the shielding effect generated by the pore when a shock wave interacts with it. The front of the pore is subjected to the full extent of the shock stress, but the pore diffracts (reflects) the wave because of the traction-free boundary condition at the pore surface, leaving the region immediately behind the pore unloaded, thus experiencing much less of the imposed loading. This phenomena is consistent with the prior observations of stress wave loading in gels \cite{Swantek2010Arrays}.

        Turning briefly to the collapsing pore itself, rather than the near-field deformation, it is worth noting that the collapsed geometry is close to an ellipsoid rather than a spheroid. This occurs, especially in this strength-dominated, low pressure regime because the shock stress is rather large compared to the lateral confining stress, leading to significantly larger deformation in the horizontal (shock) direction than the vertical (transverse) direction. An estimate for the lateral confining stress is given in \cref{eq:LatStress} using elastic theory, though this only holds near the elastic regime where minimal inelastic deformation develops. This difference becomes less severe at higher stresses (e.g., Pore--1.0) where more substantial vertical collapse begins to develop. Yet even at very high pressures (hydrodynamic regime), as reported by Escauriza, et al. for spherical pores in PMMA, the collapse in the shock direction is more drastic than the lateral directions, leading eventually to the development of jetting \cite{Escauriza2020Collapse}. 
        
        Additionally, it is possible to calculate the volume of the pore during its collapse. This is accomplished by thresholding the deformation images to produce a black and white image, from which the pore boundary can be identified using the MATLAB image processing toolbox \cite{MATLAB}. The pore boundary is then fit to two partial ellipses, for the left and right sides, as depicted in \cref{fig:PoreAsymmetry}\hyperref[fig:PoreAsymmetry]{a}. Assuming axisymmetric deformation, the volume can be computed as the sum of two ellipsoidal caps. The calculated volume evolution, normalized by the initial volume of the pore, for all four experiments is plotted in \cref{fig:LineSlice}\hyperref[fig:LineSlice]{c}. The two lower stress ($0.41$ and $0.62\,$GPa) experiments appear to reach a final collapsed volume within the first $1\,\mu$s after shock arrival, while the two higher stress experiments may still be evolving. The final measured normalized volume is plotted against the shock stress in \cref{fig:LineSlice}\hyperref[fig:LineSlice]{d}. However, because of the limited time for measurements, it is difficult to establish an exact trend for the final collapsed volume. Additional data at higher stresses would be essential to establish a relationship between shock stress and collapsed volume. The work of Escauriza, at al. on pores of a larger scale provides substantial discussion of the pore volume evolution at stresses up to $17\,$GPa, and observes that complete collapse occurs near $1.25\,$GPa \cite{Escauriza2020Collapse}. 

    \subsection{Shear Localization} \label{sec:Results-Shear}
\begin{figure*}[htpb]
            \centering
            \includegraphics[width=1.0\textwidth]{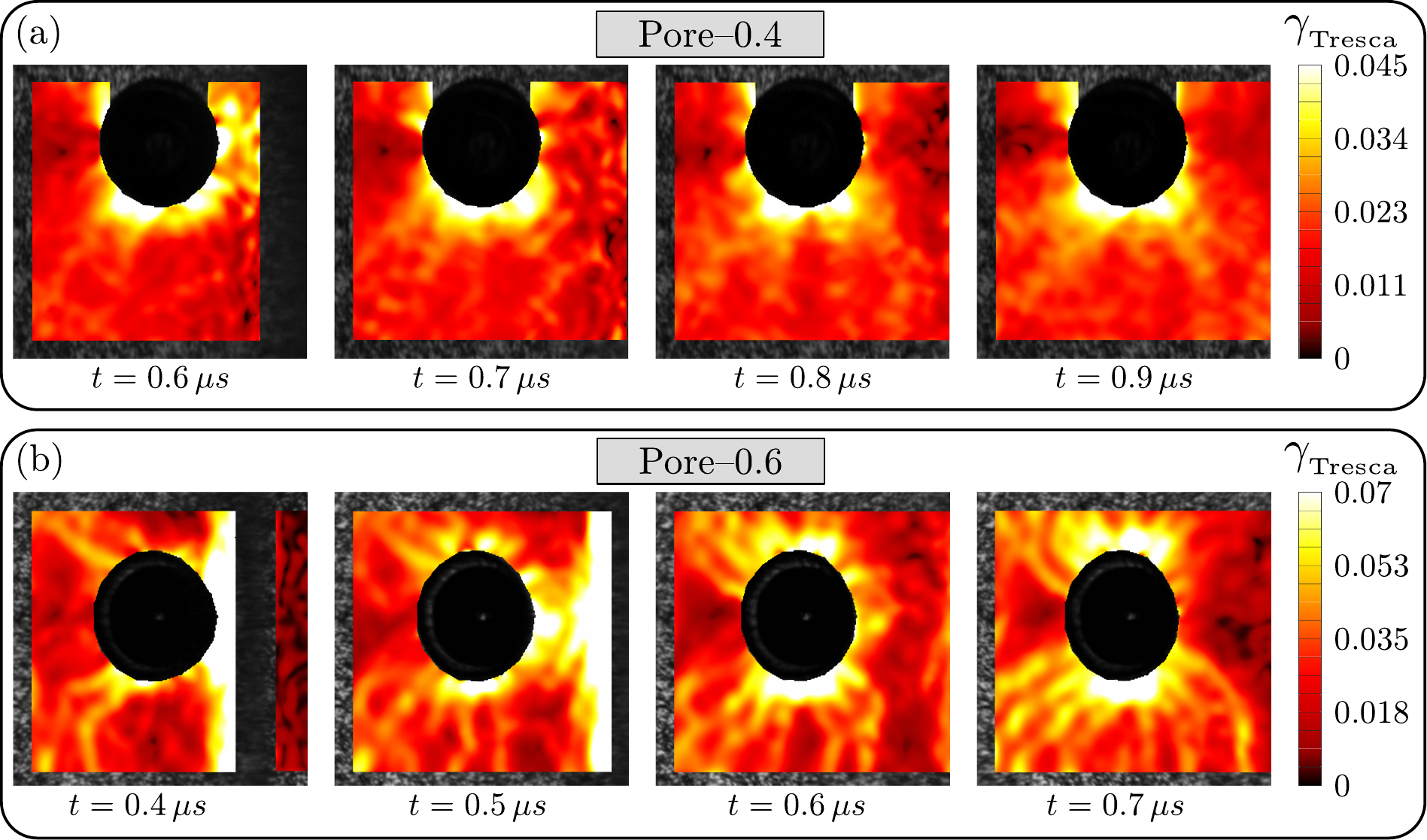}
            \caption{Tresca shear strain ($\gamma_{\text{Tresca}}$, \cref{eq:Tresca}) evolution for two lower pressure experiments after shock loading. (a) Pore--0.4 shows classical strain concentrations developed after shock. (b) Pore--0.6 evidences similar strain concentrations along with the emergence of shear bands emanating from the pore surface.}
            \label{fig:Tresca}
        \end{figure*}

        Returning to the deformation field surrounding the collapsing pore, the Tresca shear strain is computed, \begin{equation} \gamma_{\text{Tresca}} = \frac{\varepsilon_{\text{I}}-\varepsilon_{\text{III}}}{2} \label{eq:Tresca} \end{equation} which provides insights to possible localization mechanisms occurring. The symbols $\varepsilon_{\text{I}}$ and $\varepsilon_{\text{III}}$ denote the maximum and minimum principal strains. The full-field shear strain evolution is shown in \cref{fig:Tresca}. For Pore--0.4 (\cref{fig:Tresca}\hyperref[fig:Tresca]{a}), strain concentrations emerge which are reminiscent of the classical solution for stress concentrations in an infinite plate with a hole \cite{kirsch1898theorie}, indicating regions of maximum shear near the top and bottom of the pore. In the case of Pore--0.6 (\cref{fig:Tresca}\hyperref[fig:Tresca]{b}), similar strain concentrations appear after the shock passes the pore, at $t=0.4-0.5\,\mu$s. But, in addition to these concentrations, small bands begin to appear, eventually giving rise, at $t=0.7\,\mu$s, to a distinct pattern of shear bands emanating from the pore surface. The development of these bands indicates a mechanism of shear localization, via adiabatic shear banding, associated with pore collapse in PMMA. To the authors' knowledge, this work represents the first \textit{in-situ} observation of shear localization during dynamic pore collapse. 

        Because adiabatic shear bands (ASBs) are very fine structures, with a thickness in PMMA of approximately $20\,\mu$m \cite{Rittel2008DynamicFlow}, even the high magnification imaging used here is insufficient to capture the full details of these bands. Instead, using DIC, it is possible to capture the filtered deformation field. This is why the DIC measurements show bands of $\sim 6\%$ strain compared to expected strain of $\sim 100\%$ in ASBs. To confirm that the results shown here are consistent with the filtered deformation measurement of ASBs, a subset size analysis is performed. This analysis investigates the actual strain value by comparing DIC results with various correlation parameters. These correlation parameters (subset size, step size, and strain filter) influence the overall filtering influence of the DIC postprocessing, which is summarized by the virtual strain gage length \cite{Reu2015VSG}, $L_{\text{VSG}}$ (\cref{eq:VSG}). Results of the subset size analysis indicate that the bands are indeed physical features, and suggest that the actual strain magnitude is of the same order of magnitude as is expected for shear bands. Details of the analysis can be found in \cref{sec:AppendixDIC}. Recent work has also identified shear localization in hole closure experiments via post-mortem analysis for tantalum \cite{Nelms2022ClosureTantalum} and Ti-6Al-4V \cite{Lovinger2024Localization}.

    \subsection{Fracture} \label{sec:Results-Fracture}

\begin{figure*}[htpb]
        \centering
        \includegraphics[width=1.0\textwidth]{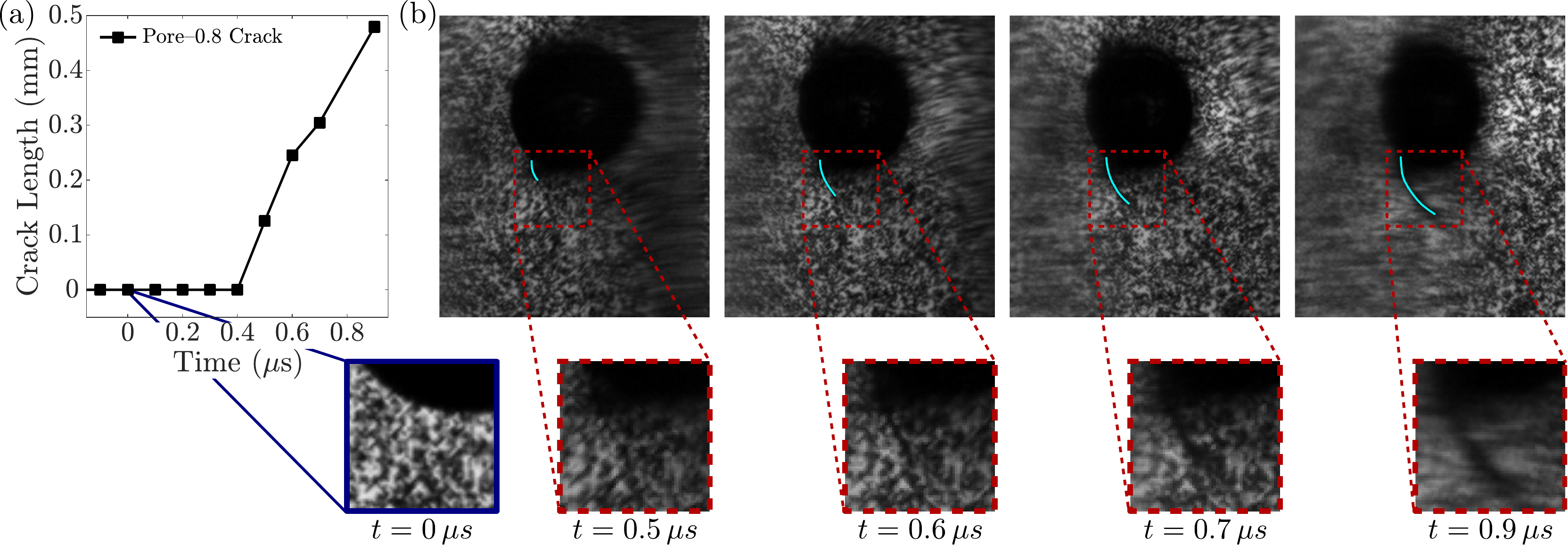}
        \caption{Time series of crack evolution in experiment Pore--0.8, with the tracked profile for each image. (a) Crack length evolution. The average crack speed is estimated to be $V_{\text{crack}} \approx 959\,$m/s. (b) Highlighted crack profile and a magnified view of the crack, provided in the inset. }
        \label{fig:FractureImages}
    \end{figure*}

    As mentioned above, Pore--0.8 reveals fracture initiation at the pore surface, beginning in the first visible image after the shock passes the pore, which can be seen in \cref{fig:FractureImages}\hyperref[fig:FractureImages]{b}. While the emergence of a crack prevents DIC analysis for this experiment, the raw deformation images enable characterization of the dynamic crack evolution during pore collapse. Escauriza, et al. have previously observed a similar fracture during pore collapse of PMMA \cite{Escauriza2020Collapse}; however, it was unclear whether boundary release waves or wave reflections from the glue interface could have applied necessary tensile loading to initiate mode I fracture. Here, the experimental design enforces nominally uniaxial strain loading conditions during the measurement window. Though the pore introduces a non-uniform stress state, theoretical and computational analyses indicate that the development of localized tension, which could drive mode I fracture, is not possible. Hence, one can conclude the crack observed here arises from shear-driven fracture, and will return to the issue in \cref{sec:Modeling,sec:Discussion}.
    
    For each image after the shock passes, the crack path is traced and displayed in \cref{fig:FractureImages}\hyperref[fig:FractureImages]{b} along with a magnified view of the crack in the inset. Additionally, the length of the crack as a function of time is shown in \cref{fig:FractureImages}\hyperref[fig:FractureImages]{a}. The average crack tip speed, $V_\text{crack} \approx 959\,$m/s is found to fall between the Rayleigh wave speed and shear wave speed of PMMA ($935$ and $1000\,$m/s, respectively \cite{Lambros1995Transonic}). However, the experimental resolution for crack speed measurement is insufficient to distinguish between crack propagation at or just above the Rayleigh wave speed. Most likely, the crack speed has approached near to the Rayleigh wave speed, consistent with the understanding that crack speeds between the Rayleigh and Shear wave speeds are unstable \cite{Broberg1960Crack, Freund1990Dynamic}. Further analysis of these results is necessary to unravel the mechanisms driving the fracture initiation point, path, as well as the modality of the fracture itself. This will be discussed in \cref{sec:Discussion}.

    The initial observations from the experimental results, presented above, indicates the intriguing presence of two failure mode transitions during pore collapse of PMMA which, until now, have not been conclusively observed \textit{in-situ}. The physics governing these transitions will be investigated by carrying out numerical simulations in \cref{sec:Modeling} along with theoretical modeling in \cref{sec:Discussion}.


\section{Modeling} \label{sec:Modeling}

    Numerical simulations were performed using the finite element software, Abaqus/Explicit \cite{Abaqus2001}, to augment our understanding of the physics involved in these experiments by providing access to information otherwise unavailable in the experimental data. 

    \subsection{Methodology}
        Simulations were set up with a 2D axisymmetric assumption, mirroring the loading conditions imposed in the experiments. Identical geometries were used for the pore size, while shrinking the target and flyer plate lateral dimensions to improve computational times. Because of the smaller lateral dimension, zero-displacement boundary conditions were applied to the lateral boundaries to maintain uniaxial strain and prevent boundary release waves. Additionally, the flyer plate and target plate geometries were created such that they shared nodes at the impact surface, thereby ensuring perfect contact and zero tilt during the simulated impact. Finally, the model was meshed with quadrilateral elements (CAX4R) in such a way that mitigates computational cost and preserves relevant physics near the pore. This was done by refining the mesh $\left(5 \, \mu\text{m element size}\right)$ near the pore interface such that the adiabatic shear bands observed in experiments could be captured, while coarsening the mesh $\left(20 \, \mu\text{m element size}\right)$ in the remainder of the model geometry. Because of the inherent length scale for localized shear features in finite element analysis (FEA), a mesh convergence study was performed. From this study, reasonable convergence was obtained with respect to the number, mode, and spacing of adiabatic shear bands for all simulations with element size at or below $5\,\mu$m. Additionally, axisymmetric simulations were performed because full 3D simulations are prohibitively expensive to perform with the converged mesh size. To confirm that the axisymmetric assumption is reasonable, 3D simulations were performed with a coarse mesh $\left(20\, \mu\text{m}\right)$ near the pore. This produced deformation and shear band patterns that were axisymmetric. Hence, 2D axisymmetric simulations are used throughout this study in place of full 3D simulations.

    \subsection{Material Models}
        To account for the thermomechanical behavior of PMMA, adiabatic heating was implemented for the simulation, along with a calibrated Johnson-Cook plasticity model \cite{Johnson1983JC} (\cref{eq:JC}) which relates the Mises yield strength $(Y)$ to the equivalent plastic strain $(\overline{\varepsilon}^{pl})$, equivalent plastic strain rate $(\dot{\overline{\varepsilon}}^{pl})$, and temperature $(T)$ through the material parameters $A$, $B$, $n$, $C$, and $m$. The model is also parameterized by the reference strain rate $(\dot{\varepsilon}_{r e f})$, reference temperature $(T_r)$, and melt temperature, $(T_m)$. The Johnson-Cook model incorporates strain rate hardening, strain hardening, and thermal softening behaviors, and has been shown to effectively capture the high strain rate material response of PMMA under confinement. Material parameters have been calibrated to fit the data of Holmquist, et al. \cite{Holmquist2016PMMA}
        
        \begin{multline}
            Y=\left[A+B \left(\overline{\varepsilon}^{pl}\right)^{n}\right]\left[1+C \ln \frac{\dot{\overline{\varepsilon}}^{pl}}{\dot{\varepsilon}_{r e f}}\right] \left[1-\left(\frac{T-T_{R}}{T_{m}-T_{R}}\right)^{m}\right]\\ \label{eq:JC}
        \end{multline}
        
        Johnson-Cook plasticity is also included for the aluminum 7075 flyer plates; however, it is worth noting that the plastic constitutive model for the flyer plate plays a minimal role in the deformation of the target plate during the window of time of interest. Instead, the standard material parameters for elasticity and wave speeds primarily govern the loading which is imposed on the target during the impact event. These quantities are all implemented in the form of shear modulus, density, and equation of state (EOS) ($U_s-u_p$ relation),
        \begin{equation} \label{eq:EOS} U_s = C_0 + Su_p \end{equation}
        where $U_s$ and $u_p$ are the shock speed and particle velocity, respectively, and $C_0$ and $S$ are material constants. \Cref{tab:Strength,tab:Properties} show all the material parameters implemented in the simulations. It is noted here that two equations of state are used for PMMA, named EOS 1 and EOS 2 to capture the non-linear profile of the shock Hugoniot data at low pressures reported by Barker \cite{Barker1970PMMA}. The simulations transition from EOS 1 to EOS 2 at the intersection point at $\sigma_{11} = 0.6\,$GPa. The EOS for Aluminum 7075 is also fit to appropriate low pressure data \cite{Marsh1980LASL}.

        \begin{table*}[ht]
            \setlength{\tabcolsep}{7.5pt}
            \captionsetup{justification=centering}
         	\caption{Parameters for the Johnson-Cook plasticity model used in the simulations.}
         	\centering
         	\footnotesize
            \begin{tabular}{ccccccccc}
            \hline\hline
            \multicolumn{1}{l}{}                & \multicolumn{8}{c}{Model parameters (\cref{eq:JC})}  \\ \cline{2-9} 
            Material                            & \begin{tabular}{@{}c@{}} $A$ \\ {[}MPa{]}\end{tabular}    & \begin{tabular}{@{}c@{}} $B$ \\ {[}MPa{]}\end{tabular}    & $n$   & $C$   & $m$   & \begin{tabular}{@{}c@{}} $\dot{\varepsilon}_{ref}$ \\ {[}1/s{]}\end{tabular} & \begin{tabular}{@{}c@{}} $T_m$ \\ {[}K{]}\end{tabular} & \begin{tabular}{@{}c@{}} $T_r$ \\ {[}K{]}\end{tabular} \\ \hline
            
            PMMA \cite{Holmquist2016PMMA}         & 210                                                     & 160                                                        & 2.95  & 0.077 & 0.74  & 1 & 398 & 298     \\
          Aluminum 7075 \cite{Brar2009Aluminum}         & 546                                                     & 678                                                       & 0.71  & 0.024 & 1.56  & 1 & 903 & 298    \\
            \hline \hline
            \end{tabular}
            \label{tab:Strength}
        \end{table*}

        \begin{table*}[ht]
 	\centering
 	\begin{threeparttable}
 		\setlength{\tabcolsep}{7.5pt}
 		\caption{Material properties and equation of state parameters.}
 		\centering
 		\footnotesize
	\begin{tabular}{cccccc}
		\hline \hline
		Material                                                              & \begin{tabular}{@{}c@{}} Density, $\rho_0$ \\ {[}kg/m$^3${]}\end{tabular} & \begin{tabular}{@{}c@{}} Specific Heat, $c$ \\ {[}J/(kg$\cdot$K){]}\end{tabular}   & \begin{tabular}{@{}c@{}} Shear Modulus, $G$ \\ {[}GPa{]}\end{tabular}            & \begin{tabular}{@{}c@{}}$C_0$ \\ {[}m/s{]}\end{tabular}           & $S$ \\ \hline
		PMMA (EOS 1) \cite{Holmquist2016PMMA,Rai2020PoreCollapse,Marsh1980LASL,Barker1970PMMA}    & 1186                                                & 1466                                                                                 & 2.19                                                                                                       & 2770                                                                & 2.11 \\
            PMMA (EOS 2) \cite{Holmquist2016PMMA,Rai2020PoreCollapse,Marsh1980LASL,Barker1970PMMA}    & 1186                                                & 1466                                                                             & 2.19                                                                                                       & 3044                                                            & 0.36 \\
		Aluminum 7075 \cite{Marsh1980LASL}                                                        & 2804                                                &  N/A                                                                                  & 26.9                                                                                                      & 5022                                                                & 1.99 \\ 
        \hline\hline                          
	\end{tabular}
	\label{tab:Properties}
 	\end{threeparttable}
        \end{table*}

    \subsection{Results}
\begin{figure*}[htpb]
            \centering
            \includegraphics[width=1.0\textwidth]{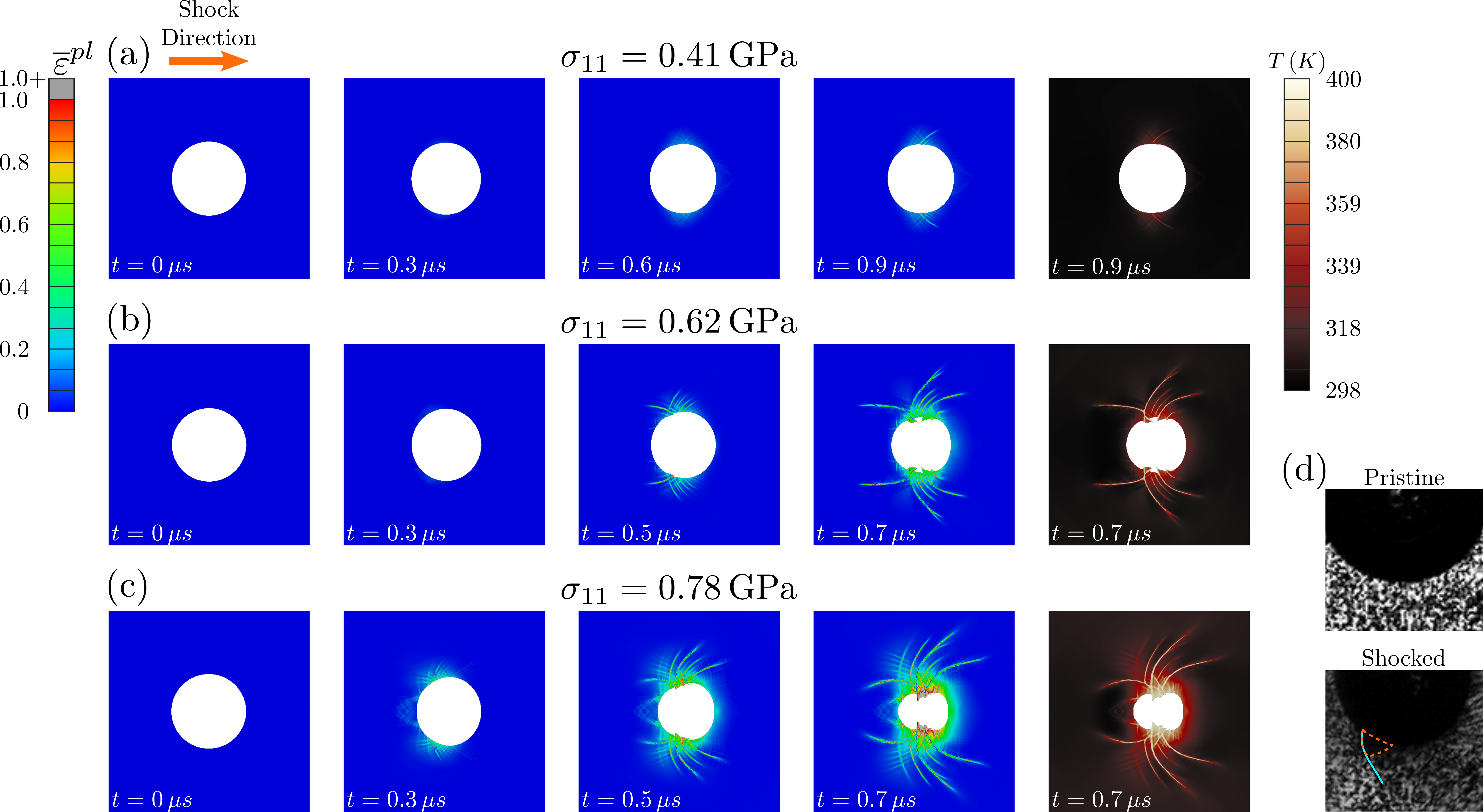}
            \caption{Results of numerical (FEA) simulations with loading conditions and measurement times consistent with experiments. (a), (b), and (c) correspond with experiments Pore--0.4, Pore--0.6, and Pore--0.8, respectively. For each simulation, the first four images depict the evolution of the pore and equivalent plastic strain, while the final image shows the temperature field at the last time instance.  (d) Highlighted view of the triangular cutout which occurs during fracture in Pore--0.8 and can be compared to (c) at $t=0.7\,\mu$s.}
            \label{fig:Simulations}
        \end{figure*}

        Results from the numerical simulations for shock stresses corresponding to the experiments are presented in \cref{fig:Simulations} with the time shifted such that the shock wave arrives at the pore surface when $t=0$. Equivalent plastic strain (shear strain) is plotted for several time instances to provide a comparison to the localized shear response observed in experiments. Upon initial observation, it is clear that the initiation of adiabatic shear bands (ASBs) is almost entirely absent for shock stress of $0.41\,$GPa (\cref{fig:Simulations}\hyperref[fig:Simulations]{a}), matching the response from experiment Pore--0.4. At $0.62\,$GPa shock stress (\cref{fig:Simulations}\hyperref[fig:Simulations]{b}), a substantial number of distributed ASBs develop. The right-most image in \cref{fig:Simulations}\hyperref[fig:Simulations]{b} reveals the maximum temperature in the shear bands to be $398\,$K which corresponds with the prescribed melting temperature in the thermal softening portion of the plasticity model. These features indicate that the first failure mode transition, from diffuse strain concentration in Pore--0.4 to localized shear via adiabatic shear banding in Pore--0.6, is governed simply through thermo-viscoplastic mechanics in the form of the Johnson-Cook model, which was implemented in FEA and replicated the transition. Given that the target material, PMMA, is amorphous, one would expect this to be the case, as the development of ASBs in PMMA cannot be affected by microstructure as has been suggested for polycrystalline metals \cite{Rittel2008Recrystallization}. 
        
        Finally, the model at $0.78\,$GPa shock stress (\cref{fig:Simulations}\hyperref[fig:Simulations]{c}) evolves toward large deformation, growth of many shear bands, and development of what appears to be a dominant shear band with a very similar initiation point and trajectory as the crack observed in Pore--0.8. This dominant shear band suggests that the subsequent fracture observed experimentally is indeed shear-driven and occurs along the ASB path---enabled by the weakened material state which was realized after undergoing significant shear deformation \cite{Bai1992ASB}. The FEA model also reveals a triangular feature resulting from the intersection of two ASBs in \cref{fig:Simulations}\hyperref[fig:Simulations]{c} at $t=0.75\,\mu$s. This feature is very similar in location and geometry to that observed experimentally in Pore--0.8 where the crack initiates, which is highlighted in \cref{fig:Simulations}\hyperref[fig;Simulations]{d}. Again, this reinforces the ability of the FEA to capture the dominant ASB modes which lead to shear-driven fracture in the experiments. This fracture of a triangular chip has also been recently observed by Lovinger, et al. \cite{Lovinger2024Localization} in cylindrical hole closure via post-mortem analysis.

        Regarding the actual distribution and curvature of the ASBs in the $0.62\,$GPa simulation, it is not clear why the patterns fail to match the experiments. One possible reason is that at a given material location, two orthogonal directions of maximum shear exist, but the mesh may give slight preference toward one direction over the other, leading to the predominant occurrence of shear bands which curve toward the shock loaded direction. While the exact ASB distribution for Pore--0.6 is not captured, and the FEA implementation used here does not attempt to incorporate fracture to compare directly with Pore--0.8, the qualitative replication of failure modes provides helpful insight to the physics of the failure mode transitions. Most notably, the simulations confirm that the transition to adiabatic shear banding is governed by thermo-viscoplastic mechanics, and reinforces the idea that the eventual failure via fracture is indeed shear-driven and is enabled by the weakened material along the dominant ASB. This discussion will be continued along with further investigation through theoretical modeling next, in \cref{sec:Discussion}, with the goal of understanding the ASB trajectories and driving mechanisms for fracture.

\section{Discussion} \label{sec:Discussion}
    The experiments conducted in this study, along with numerical simulations, have uncovered a distinct transition in failure modes with increasing shock stress. First, shear localization develops via many distributed adiabatic shear bands, and second, fracture occurs along a dominant adiabatic shear band which is enabled by material softening in the shear band. In these experiments on a single, $800\,\mu$m diameter pore in PMMA, the first transition was found to occur between $0.41$ and $0.62\,$GPa shock stress, while the second is between $0.62$ and $0.78\,$GPa. However, these transitions are certain to be material dependent, and may have additional dependence on (i) the pore shape and pore size, and (ii) the configuration of an array of pores. With the transitions observed in experiments and supported by simulations, one can now turn to the topic of the driving mechanisms for failure mode transition, initiation locations, and propagation/arrest. 

    \subsection{Failure Mode Transitions}
\begin{figure*}[htpb]
            \centering
            \includegraphics[width=0.8\textwidth]{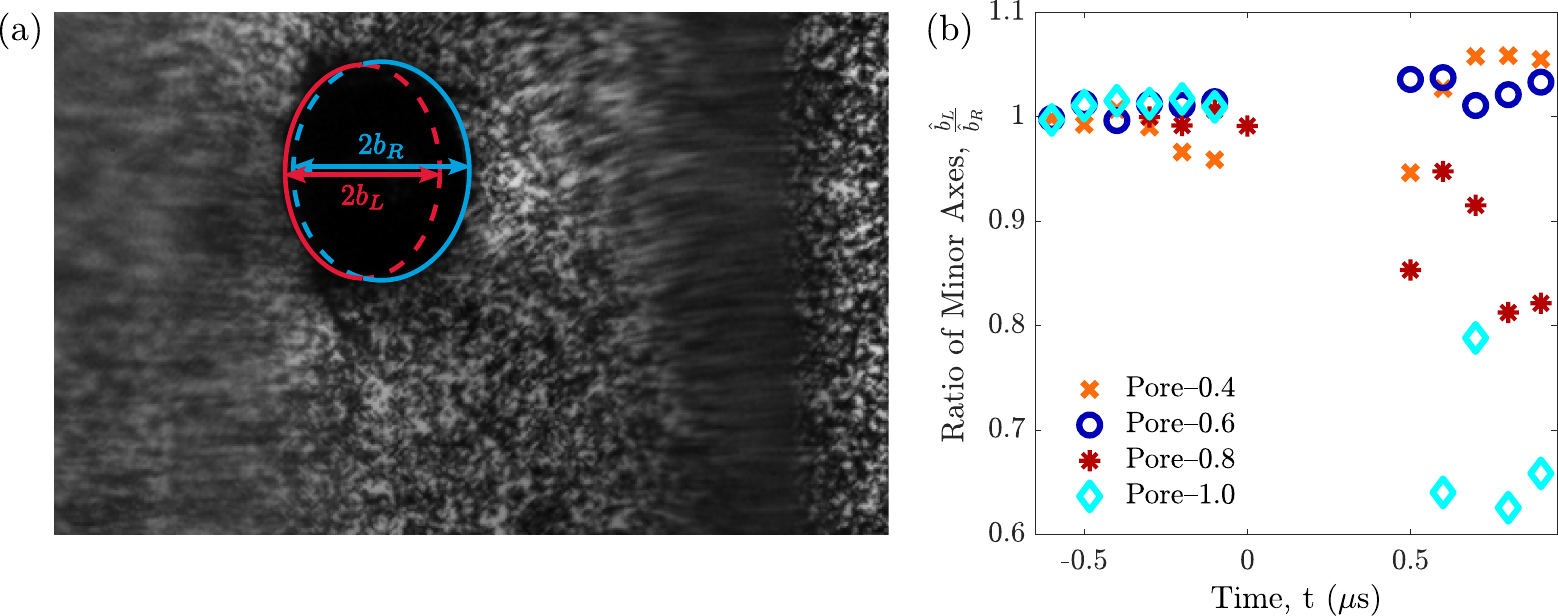}
            \caption{Pore collapse asymmetric shape analysis. (a) Example of partial ellipses fit to the collapsing pore to characterize volume and asymmetry of pore shape. $b_L$ and $b_R$ refer to the minor axis length for the left (shock impinging face, red) and right (blue) sides of the pore, respectively. The solid curves correspond to the actual partial ellipse fits to the pore outline, while the dashed curves depict the remaining portion of the ellipses which do not fit the pore outline. (b) The normalized ratio of the minor axes of the left and right ellipses, $\hat{b}_L/\hat{b}_R$, where $\hat{b} = b/b^{0}$, $b^{0}_{L}$ and $b^{0}_{R}$ are the minor axes lengths of the initial pore left and right sides respectively; $b^{0}_{L} \approx b^{0}_{R} \approx R$ (nominal pore radius).}
            \label{fig:PoreAsymmetry}
        \end{figure*}

        As has been discussed previously, it is clear that the transition to failure via adiabatic shear banding is governed by thermo-viscoplastic mechanics. In particular, even at relatively low shock stresses, the presence of a pore creates large stress concentrations, inducing significant plastic deformation leading to inelastic (plastic) heating and thermal softening \cite{DoddBai2012ASB}. These conditions are ideal for the development of adiabatic shear bands. However, the physics governing the transition from distributed ASBs to the dominance of a single ASB and subsequent fracture is less obvious. Toward understanding this transition, the symmetry of the pore collapse is analyzed, having visually observed a substantial asymmetry in the raw deformation images for Pore--0.8 (\cref{fig:DeformationImages}\hyperref[fig:DeformationImages]{c}). This is accomplished by fitting a partial ellipse to each side of the pore, as was described in \cref{sec:Results-Deformation} and shown in \cref{fig:PoreAsymmetry}\hyperref[fig:PoreAsymmetry]{a} for one image from Pore--0.8. Plotting the ratio of minor axes in \cref{fig:PoreAsymmetry}\hyperref[fig:PoreAsymmetry]{b}, a clear difference emerges between the lower pressure (Pore--0.4 and Pore--0.6) and higher pressure (Pore--0.8 and Pore--1.0) experiments. The lower pressure experiments exhibit fairly symmetric collapse, marked by a ratio of minor axes of the two ellipses (left and right) close to 1. But the higher pressure experiments display substantial asymmetry, as the left side (on which the shock impinges first) compresses more than the right (which is shielded), leading to a ratio well below 1. It is noteworthy that the ratio does fluctuate, first dipping significantly below 1 in the first frame after the shock has passed the pore (the first data point after losing view of the pore), then rising as the right side compresses after being shock loaded, and finally settling back down as the pore reaches its final steady state shape. This asymmetric collapse behavior is another unique feature of shock compression of pores, compared to the classical symmetric theories \cite{Herrmann1969Constitutive,Carroll1972Relation}, static compression, or other transient loadings \cite{McGhee2023Microcavitation}. This large asymmetry coincides directly with the initiation of fracture in Pore--0.8, which nucleates and propagates a very short distance at $t=0.5\,\mu$s as shown in \cref{fig:FractureImages}\hyperref[fig:FractureImages]{a}, corresponding to the time at which the axes ratio first is measured to be well below 1. 

        The asymmetry is driven in part by the transient nature of shock loading, in which the left side of the pore is deformed before the right side is fully loaded, leading inherently to a time shift in the amount of deformation on the right side compared to the left. These transients also would drive the failure initiation location preferentially to the left, rather than the center where the maximum shear should develop in quasistatic loading. However, an additional mechanism, and the reason for which the shocked pore remains asymmetric even after reaching a steady state, are the kinematics imposed by wave interactions with the pore surface.  In particular, upon arrival at the pore surface, the shock wave propagates forward, diffracting around the pore and leading to significant longitudinal stress at the top and bottom of the pore. However, at the front of the pore (left side), the shock wave reflects off the pore owing to the free surface (traction-free boundary condition). Associated with this release, and consistent with classical shock physics \cite{Davison2008Shock}, the particle velocity at the front surface of the pore could be doubled compared to that of the top or bottom. This phenomena arising from wave interactions is the same one by which jets develop at much higher shock stresses, where momentum carries the front surface forward and impinges on the back surface of the pore. Along the surface of the pore between the front and top/bottom, where the surface is neither parallel nor perpendicular to the shock loading direction, a multiaxial response is generated with a partial release. In other material systems, such as copper \cite{Lind2021Closure}, these kinematics are accommodated through large plastic deformations. Even in PMMA, in the hydrodynamic regime, it is known that spherical pores do eventually develop jets \cite{Escauriza2020Collapse} at sufficiently high stresses, likely enabled by shock heating which melts the PMMA. In the most extreme case of the complete absence of strength in the matrix material (i.e., fluids), this phenomena has been well studied and clearly described through the shock bubble interaction problem \cite{markstein1957shock,rudinger1958shock,haas1987interaction}. Here in the strength-dominated regime, however, because of the more brittle nature of PMMA, the kinematic frustration generates large concentrations of shear strain. This leads to shear localization and eventual fracture, instead of developing more diffuse plastic deformation and uniform softening. 

        Understanding the influence of asymmetric collapse resulting from shock compression helps to explain the transition from distributed ASBs in Pore--0.6 to dominant ASB and shear fracture in Pore--0.8. Interestingly, Pore--1.0 also shows substantial asymmetry, indicating that shear fracture could be present; however, the experimental images are occluded, preventing any conclusive claims. Lovinger, et al. observed post-mortem shear fracture in Ti-6Al-4V \cite{Lovinger2024Localization} during hole closure at various shock stresses, indicating that this phenomena is more widespread and not limited only to PMMA or a very specific shock stress. However, in PMMA it is also possible that the brittle-to-ductile transition which occurs under significant confinement \cite{Rittel2008DynamicFlow} may mitigate the fracture response at higher shock stresses. Coupled with material softening resulting from shock heating at higher shock stresses and the possible suppression of crack propagation via confining stress, another transition from shear fracture may be possible beyond $0.8\,$GPa shock stress. Finally, it is noteworthy that this fracture behavior could be highly dependent on the spatial scale. The transient loading mentioned above should become negligible for very small pores, and shear localization may become irrelevant when the pore scale approaches the inherent length-scale of ASBs in PMMA ($20\,\mu$m \cite{Rittel2008DynamicFlow}). Alternatively, at large pore scales (e.g., engineered structures and metamaterials), the effect would likely become more pronounced.


    \subsection{Adiabatic Shear Band Spacing and Paths}
\begin{figure*}[htpb]
            \centering
            \includegraphics[width=1.0\linewidth]{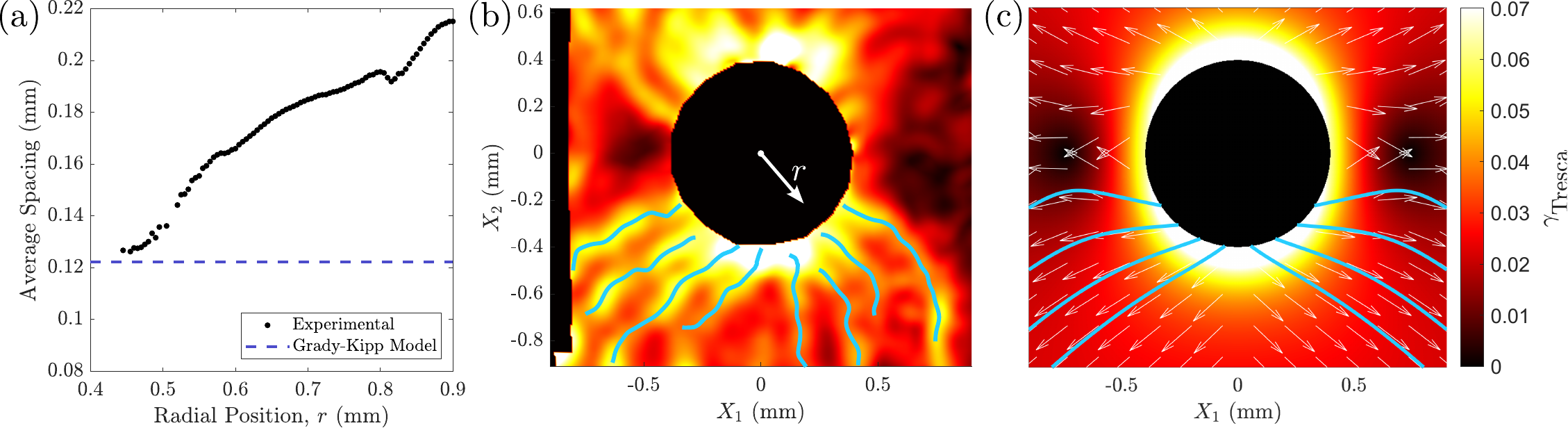}
            \caption{Comparison of shear band spacing and paths: experiments and theory. (a) Experimental data for the spacing between tracked, developed ASBs in Pore--0.6 are compared to the theoretical prediction of Grady and Kipp \cite{Grady1987ASBSpace}. (b) DIC results for Tresca shear strain, $\gamma_{\text{Tresca}}$, in experiment Pore--0.6 at $t=0.7\,\mu$s with traced ASB paths outlined in light blue. Results are shown in the reference (undeformed) configuration. (c) Elastic solution for Tresca shear strain, $\gamma_{\text{Tresca}}$. Directions of maximum shear indicated by white arrows. Light blue curves represent streamlines along the path of maximum shear, beginning at the estimated initiation sites for ASBs in experiment Pore--0.6. }
            \label{fig:SpacingPathsASB}
        \end{figure*}

        The next question to address is that of shear band spacing and paths, which should provide insight to the physics behind ASB distribution and self-organization which is observed in Pore--0.6. To this end, the shear bands are tracked in the reference (Lagrangian or undeformed) configuration. This is done by manually estimating one point along each shear band from the DIC data, computing the location of the nearest local maximum shear strain, and tracking the local maxima (shear strain peaks) along the length of the shear bands. The final traced shear bands are overlaid on the Tresca shear strain (\cref{eq:Tresca}) plot in \cref{fig:SpacingPathsASB}\hyperref[fig:SpacingPathsASB]{b}. From there, one can calculate the spacing between each shear band and its neighbors as a function of radial position. These results are presented in \cref{fig:SpacingPathsASB}\hyperref[fig:SpacingPathsASB]{a} along with a comparison to theory. 

        Following a similar approach as in previous studies \cite{Nesterenko1998SelfOrganization,Xue2002SelfOrganization,Xue2003ThickWalled,Ramesh2006Formation}, one calculates the theoretical spacing using the Grady-Kipp model \cite{Grady1987ASBSpace}. During deformation, as adiabatic shear bands form, they weaken and flow to very large strains, resulting in an unloading process of the neighboring material. This is the notion which Grady and Kipp capture through a momentum diffusion model, determining the spacing based on the speed of the unloading front. The spacing prediction which results from their analysis \cite{Grady1987ASBSpace} has been summarized by Nesterenko, et al. \cite{Nesterenko1998SelfOrganization} as follows: \begin{equation} L_{\text{GK}} = 2\left(\frac{9kc}{\dot{\gamma}^{3}a^{2}\tau_{0}}\right)^{1/4} \label{eq:GK} \end{equation} where $k$ and $c$ are the thermal conductivity and specific heat capacity, respectively. The applied shear strain rate is $\dot{\gamma}$, the flow stress at room temperature is denoted $\tau_0$, and the temperature dependence of the flow stress is characterized through the linear thermal softening parameter, $a$.

        \begin{equation} \tau = \tau_0[1-a(T-T_0)] \label{eq:LinThermSoft} \end{equation}

        Note that the flow stress is defined here as $\tau_0 = \sigma_0/2$, where $\sigma_0$ is the yield stress at room temperature under uniaxial stress loading. All the relevant parameters have been fit to the same dataset used in the plasticity model for the numerical simulations \cite{Holmquist2016PMMA}, and are summarized in \cref{tab:GKModel}. The applied strain rate is estimated based on shear strain rate near the pore surface from a numerical simulation of pore collapse at $0.62\,$GPa shock stress without thermal softening included in the analysis.

        \begin{table*}[ht]
            \setlength{\tabcolsep}{7.5pt}
            \captionsetup{justification=centering}
         	\caption{Parameters for Grady-Kipp adiabatic shear band spacing model.}
         	\centering
         	\footnotesize
            \begin{tabular}{ccccccc}
            \hline\hline
            Material                                                 & $k$ [W/(m$\cdot$K)]       & $c$ [J/(kg$\cdot$K)]          & $\dot{\gamma}$ [s$^{-1}$]     & $\tau_0$ [MPa]        & $T_0$ [K]     & $a$ [K$^{-1}$]    \\ \hline
            PMMA \cite{Rai2020PoreCollapse,Holmquist2016PMMA,Assael2005Thermal}        & 0.19                    & 1466                        & 0.45$\times$10$^6$            & 53                    & 298           & 0.0061            \\ 
            \hline \hline
            \end{tabular}
            \label{tab:GKModel}
        \end{table*}

        Comparing the experimental results from Pore--0.6 to the Grady-Kipp model, one finds close agreement near the pore surface. This suggests that the momentum diffusion mechanism generally captures the physics which govern the spacing of ASBs at their most densely packed location (e.g., the pore surface) during symmetric collapse. This conclusion seems reasonable, at least for the analysis of ASBs which initiate, propagate outward from the pore surface, and attain sufficient strain magnitude such that they are captured via DIC. It remains possible that other ASBs initiate near the pore surface, but the loading and unloading mechanism allows some ASBs to fully develop while others die out. Such existence of very small ASBs which are unable to compete with the growth of neighboring ASBs is well known \cite{Xue2002SelfOrganization}. A numerical study \cite{Ramesh2006Formation} by Zhou, et al. distinguished between initiated ASBs and developed ASBs, finding the Grady-Kipp model to predict the spacing of developed shear bands, while other models \cite{wright1996scaling,molinari1997collective} performed better when considering the spacing between all initiated shear bands.  Finally, considering the Pore--0.8 GPa experiment, one dominant ASB develops and fracture occurs along this shear band (\cref{fig:FractureImages,fig:Simulations}\hyperref[fig:Simulations]{c}). However, this does not preclude other ASBs from developing. This cannot be captured in experiment because the fracture prevents DIC analysis of the remaining deformation field. Still, one would expect that fracture would serve to unload the material even more effectively than shear bands, leading to the conclusion that other ASBs could arrest in the presence of a dominant shear fracture, such as is seen in Pore--0.8.

        Having considered throughout this section the topics of ASB and crack initiation as well as the spacing between shear band initiation sites, one can now turn to the topic of shear band paths. Recalling the elastic solution introduced along with the line slices in \cref{fig:LineSlice}, the full field solution is considered, with a particular interest in the directions of maximum shear which would direct the shear band paths. Details of the elastic solution are provided in \cref{sec:AppendixElastic}. \Cref{fig:SpacingPathsASB}\hyperref[fig:SpacingPathsASB]{b-d} summarizes the relevant results, first reproducing the representative experimental image of distributed shear bands in Pore--0.6 (\cref{fig:SpacingPathsASB}\hyperref[fig:SpacingPathsASB]{b}). Then the Tresca shear strain (\cref{eq:Tresca}) deduced from the elastic solution is plotted in \cref{fig:SpacingPathsASB}\hyperref[fig:SpacingPathsASB]{c} for a far field loading, $\sigma_0=0.62$ GPa, along with corresponding lateral confining stress superposed. Additionally, the direction of maximum shear, the bisection of the first and third principal directions, is indicated by white arrows.

        Qualitative agreement between the dynamic experimental and elastostatic theoretical Tresca strain field is clear to visualize, with major concentrations of similar magnitude on the top and bottom of the pore. The primary interest, however, is in the directions of maximum shear. To compare with experimental traces of the shear band patterns, the average angular spacing between the eight bands at the pore surface is calculated and found to be 16 degrees. Thus, initiation points are estimated to be evenly spaced 16 degrees apart, and symmetric across the $X_2$ axis. Streamlines are computed, initiating at these sites and following the path of maximum shear, and are plotted as blue overlaid curves. While not a perfect match to the experiments, it is interesting to see elastic theory capture the qualitative response of such complex phenomenon. Considering the differences in the initial shock compressed loading state to this static elastic solution, as well as the dynamic evolution involving the development of shear bands and subsequent unloading of the neighboring regions, it is expected that the actual response will not follow the elastic solution exactly. Still, it offers a good comparison and insights for the localization behavior of pores under extreme loading.

    \subsection{Crack Path and Arrest}
        
        \begin{figure*}[htpb]
            \centering
            \includegraphics[width=0.85\textwidth]{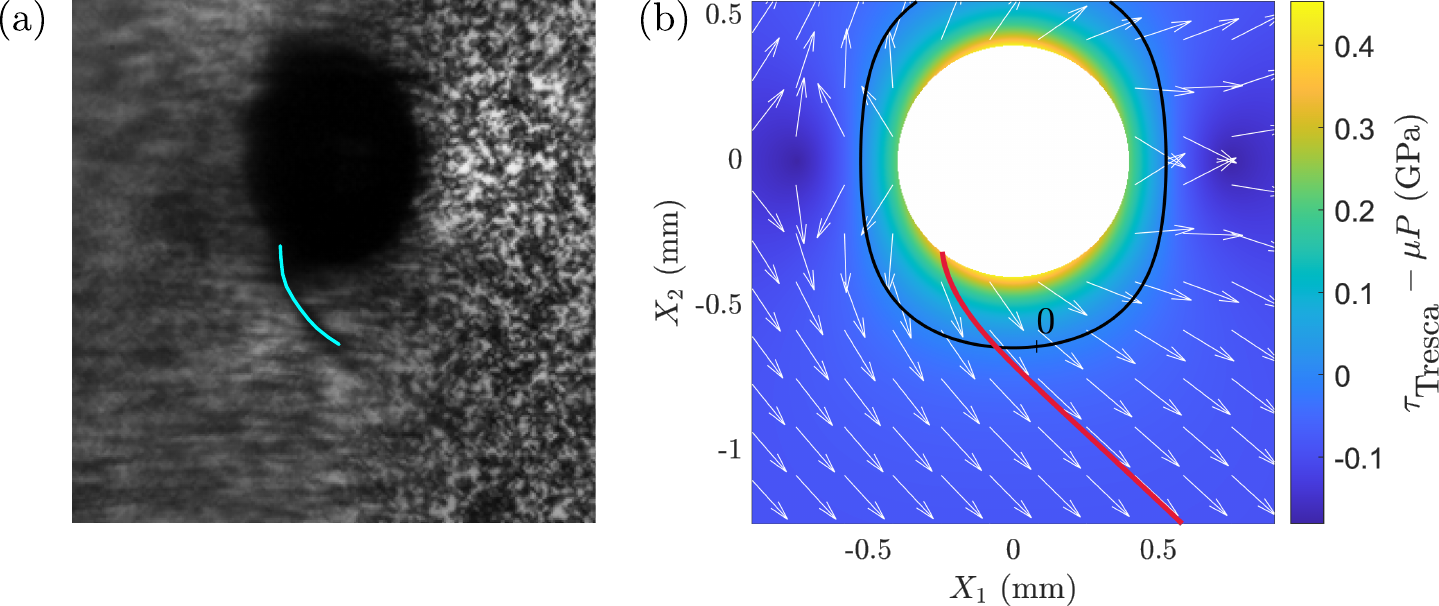}
            \caption{Comparison of experimental crack path in Pore--0.8 to the direction of maximum shear predicted by theory. (a) Raw deformation image from experiment Pore--0.8 at $t=0.9\,\mu$s with shear fracture highlighted in light blue. (b) Elastic solution for driving force: $\tau_d = \tau_{\text{Tresca}}-\mu P$ (\cref{eq:DrivingForce}) with direction of maximum shear (white arrows) and streamline (red) beginning at the fracture initiation location. Also shown is the contour (black) where the driving force is zero.}
            \label{fig:FracturePathArrest}
        \end{figure*}

        To investigate the crack path and arrest, a similar procedure, which uses the elastic solution to determine the direction of maximum shear, is undertaken for the 0.78 GPa case (Pore--0.8), the results from which are shown in \cref{fig:FracturePathArrest}. Estimating the initiation point is more straightforward for Pore--0.6, because the DIC results allow analysis in the reference configuration. Instead, for Pore--0.8, an ellipse is fit to the pore in the deformed state, and the fracture initiation point is manually identified on the fitted ellipse. From there, the angular location along the ellipse is calculated and applied to the circular contour in the reference configuration. Next, the streamline is generated to follow the path of maximum shear. One can imagine based on the results in \cref{fig:SpacingPathsASB}\hyperref[fig:SpacingPathsASB]{c} that the streamline would not resemble the crack path. However, recalling the definition for direction of maximum shear and the fact that principal directions are independent of sign, there exists an alternative set of maximum shear directions which is orthogonal to the first set. Computing the streamline with this alternative set of maximum shear vectors, the result in \cref{fig:FracturePathArrest}\hyperref[fig:FracturePathArrest]{b} is achieved, which corresponds nearly identically to the fracture path observed in Pore--0.8 (\cref{fig:FracturePathArrest}\hyperref[fig:FracturePathArrest]{a}). This is another clear indication that fracture indeed occurs in shear and along the path of a shear band. Furthermore, one can consider the driving force for crack propagation, and compare it with the final crack tip location at $t=0.9\,\mu$s. At this time, it appears the crack tip has arrested, as the subsequent images show no evidence of further propagation. The work by Lovinger, et al. \cite{Lovinger2024Localization} also indicates crack arrest in Ti-6Al-4V for several experiments, though at higher pressures the crack propagates through to the boundary of the specimen. A simplistic shear driving force ($\tau_{d}$) on the crack is considered: 
        \begin{equation} \tau_d = \tau_{\text{Tresca}}-\mu P  \label{eq:DrivingForce} \end{equation} 
        where $\mu$ is the coefficient of friction, taken to be 0.3 \cite{Bouissou1998PMMA}, $\tau_{\text{Tresca}}$ is the Tresca shear stress defined in \cref{eq:TrescaStress}, and $P$ is the pressure (2D), or compressive loading normal to the path of maximum shear, defined in \cref{eq:Pressure}. Compressive stress is taken to be negative, $\sigma_{\text{I}}$ is the largest principal stress, and $\sigma_{\text{III}}$ is the smallest principal stress.  Note that compressive pressure ($P$) is taken to be positive here.

        \begin{equation} \tau_{\text{Tresca}} = \frac{\sigma_{\text{I}} - \sigma_{\text{III}}}{2} \label{eq:TrescaStress} \end{equation}
        \begin{equation} P = \frac{-\left(\sigma_{I} + \sigma_{III}\right)}{2}  \label{eq:Pressure} \end{equation} 
        
        Such a failure criterion (\cref{eq:DrivingForce}) has been considered before in numerical models for fracture of brittle materials \cite{Camacho1996Brittle} and to understand failure of brittle materials under confinement \cite{Chen2000Brittle}. It is assumed that when the driving force goes to zero, the shear crack would stop propagating forward. This driving force is plotted as the color map in \cref{fig:FracturePathArrest}\hyperref[fig:FracturePathArrest]{b}, along with a black contour curve where the driving force is zero. Hence, the shear fracture arrest can be predicted to occur near the intersection of the streamline and the zero-driving force contour. As it turns out, this location coincides very closely with the experimental result of Pore--0.8. This finding indicates that the driving force (\cref{eq:DrivingForce}) is indeed a good estimate for shear fracture behavior in these experiments. It also clarifies the mechanism for the crack arrest observed in Pore--0.8, and possibly in the recent results on hole closure in Ti-6Al-4V \cite{Lovinger2024Localization}.

\section{Conclusion} \label{sec:Conclusion}
    
    In summary, plate impact pore collapse experiments were conducted at shock stresses of $0.4-1.0\,$GPa, using the recently developed internal DIC technique to perform quantitative measurements of deformation in shock compression experiments \cite{Lawlor2024IntDIC}. These experiments led to the first \textit{in-situ} observation of shear localization during pore collapse via adiabatic shear banding, and also confirmed the previous \textit{in-situ} \cite{Escauriza2020Collapse} and post-mortem \cite{Lovinger2024Localization} observations of crack nucleation from the pore surface. From these insights, two failure mode transitions were observed as the shock stress increased: first, from diffuse strain concentration to failure via adiabatic shear localization, and second, to dynamic fracture. Numerical simulations demonstrated that thermo-viscoplastic modeling qualitatively captures these failure mode transitions. Further, they confirmed the fracture at $0.78\,$GPa to be shear-driven fracture (mode-II) and indicated the development of a dominant shear band at high pressures which enabled fracture through the weakened material inside the shear band. 
    
    Analysis of the pore asymmetry evolution during collapse demonstrated a correlation between large asymmetry in the collapsed pore shape (which arises from wave interactions with the pore) and nucleation of dynamic fracture. It is proposed that the wave interactions and subsequent asymmetric collapse create shear strain concentrations which lead to the development of a dominant ASB. The ASB effects a partial stress (energy) release, and eventually gives way to shear fracture that provides an additional stress (energy) release. It also accommodates the large deformations imposed by the shock wave interactions with the pore and surrounding material. The physics governing the distribution of ASBs at $0.62\,$GPa are clarified through comparison of ASB spacing with the theoretical model proposed by Grady and Kipp \cite{Grady1987ASBSpace}, which fits well to the experimentally measured spacing of developed shear bands at the pore surface. Thus, the fundamental mechanism in the model: unloading of nearby material through the development of ASBs, which is captured through a momentum-diffusion model \cite{Grady1987ASBSpace}, should also govern the number and spacing of developed shear bands at the pore surface. 
    
    Static elastic theory helped elucidate the physics which determine the paths and arrest of ASBs and cracks. The direction of maximum shear is found to effectively replicate the paths which the ASBs and the crack follow, providing a simple and reasonable method for predicting and understanding the failure paths during pore collapse. Finally, using a simple estimate (\cref{eq:DrivingForce}) for the driving force for shear fracture, the arrest location of the crack tip can be accurately replicated.

    Future work will aspire to extend the fundamental understanding of the deformation and failure for heterogeneous materials beyond a single pore. Extending the experimental technique, and leveraging the insights gained in this work, one could investigate the interactions between multiple pores in various configurations. Additionally, the influence of pore size on the pore collapse phenomenon and its associated deformations and failure modes, is likely to yield intriguing results. Synthesis of these types of experiments and implementation in multiscale modeling may also greatly enhance our understanding of the underlying mechanisms for the continuum response of porous materials. Additionally, understanding the role of hard inclusions---the fundamental building block of particulate composites---on the neighboring matrix material is of similar interest. Finally, implementing the technique presented here with phase contrast imaging at a synchrotron x-ray source could minimize the issues associated with optical distortions, enable investigation of longer loading periods and higher stresses, and enlarge the list of material candidates for study.

\begin{acknowledgments}
The research reported here was supported by the DOE/NNSA (Award No. DE-NA0003957), which is gratefully acknowledged. The authors acknowledge the Army Research Laboratory (Cooperative Agreement Number W911NF-12-2-0022) for the acquisition of the high speed cameras.
\end{acknowledgments}


\section*{\large{Author Declarations}}

\subsection*{Conflict of Interest}
The authors have no conflicts to disclose.

\subsection*{Author Contributions}
\noindent\textbf{Barry Lawlor:} Conceptualization (lead); Methodology (lead); Investigation (lead); Formal analysis (lead); Visualization (lead); Writing – original draft (lead); Writing – review and editing (equal)\\
\textbf{Vatsa Gandhi:} Conceptualization (supporting); Methodology (supporting); Writing – review and editing (equal) \\
\textbf{Guruswami Ravichandran:} Supervision (lead); Funding acquisition (lead); Conceptualization (supporting); Formal analysis (supporting); Writing – review and editing (equal). \\

\subsection*{Data Availability Statement}

The data that support the findings of this study are available from the corresponding author upon reasonable request.

\begin{appendices}
\appendixtitleon
\section{Elastic solution for a pore subjected to multiaxial loading} \label{sec:AppendixElastic}

        \begin{figure*}[htpb]
            \centering
            \includegraphics[width=0.8\textwidth]{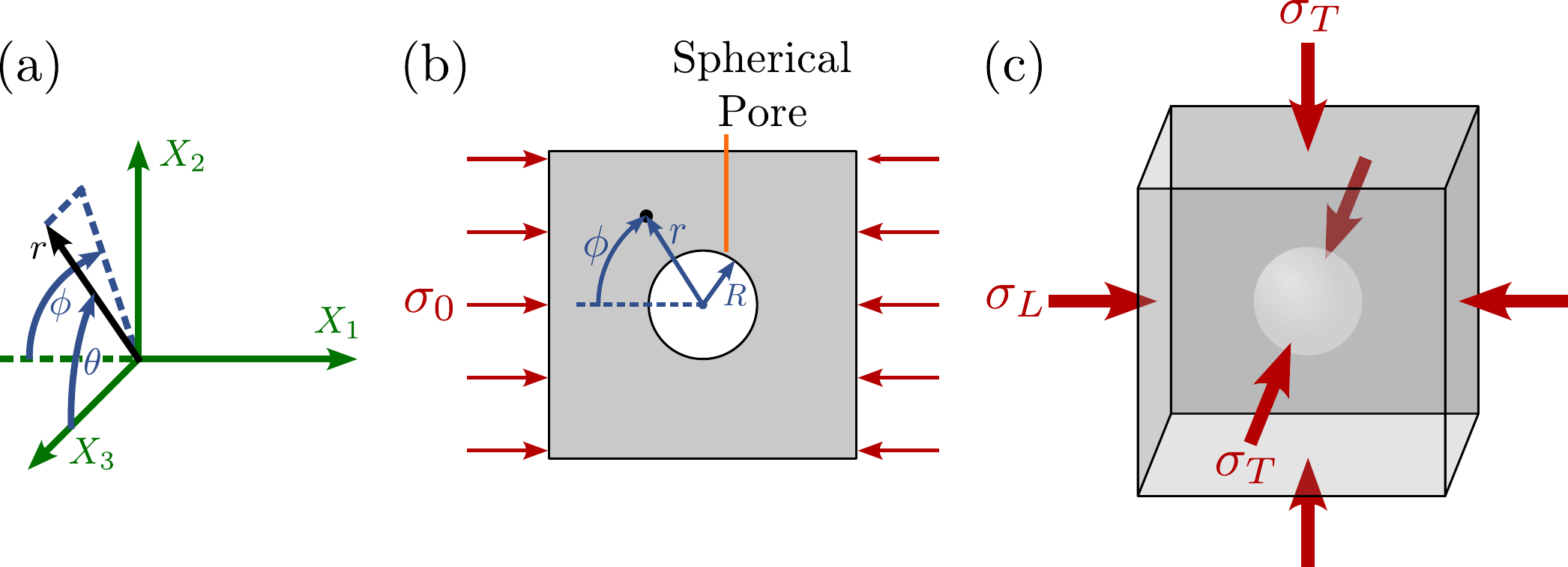}
            \caption{Schematic for elastic solution of a spherical pore in an infinite body. (a) Spherical and Cartesian coordinate systems overlaid. (b) 2D diagram of uniaxial stress ($\sigma_{11}=\sigma_{0}$) static loading on body with a spherical pore. 2D view is shown at the mid-plane ($X_3=0$, $\theta=\pi /2$). (c) Superposition of shock (longitudinal) stress, $\sigma_{11}=\sigma_L$, and lateral (transverse) confining stress, $\sigma_{22}=\sigma_{33}=\sigma_T$.}
            \label{fig:ElasticDiagram}
        \end{figure*}
        
        The general solution for static, uniaxial stress loading of an infinite, isotropic, linear elastic body with a spherical pore inside, derived by Southwell, \cite{Southwell1926Concentration} is shown in \cref{eq:s_rr_gen,eq:s_tt_gen,eq:s_pp_gen,eq:s_rp_gen}. The notation is adapted to suit the convention for this work, and uses the spherical coordinate system (centered at the pore center) depicted in \cref{fig:ElasticDiagram}\hyperref[fig:ElasticDiagram]{a}, where r is the position along the radial coordinate, $\theta$ is the polar angle, and $\phi$ is the azimuthal angle. \Cref{fig:ElasticDiagram}\hyperref[fig:ElasticDiagram]{b} shows a two-dimensional view at the plane of interest ($X_3=0$ or $\theta=\pi /2$; comparable to the experimental speckled mid-plane), with an arbitrary applied longitudinal stress, $\sigma_{11} = \sigma_{0}$. Note that R represents the nominal initial pore radius. Values for the elastic parameters (Elastic modulus, $E = 5.76\,\text{GPa}$, and Poisson's ratio, $\nu = 0.42$) are taken for PMMA under dynamic loading with confinement \cite{Rittel2008DynamicFlow}.

        \begin{widetext}
            \begin{equation} \sigma_{rr} = \sigma_{0}\left\{ \text{cos}^2 \phi + \frac{1}{14-10\nu}\frac{R^3}{r^3} \left[-38 + 10\nu + 24\frac{R^2}{r^2} + \left(50 - 10\nu - 36\frac{R^2}{r^2}\right)\text{sin}^2\phi\right] \right\} = \sigma_{0} K_{rr}\left(\phi,r\right) \label{eq:s_rr_gen} \end{equation}

            \begin{equation} \sigma_{\theta \theta} = \frac{\sigma_{0}}{2\left(7-5\nu\right)} \frac{R^3}{r^3} \left[9 - 15\nu -12\frac{R^2}{r^2}-15\left(1-2\nu - \frac{R^2}{r^2}\right)\text{sin}^2\phi\right] = \sigma_{0} K_{\theta \theta}\left(\phi,r\right) \label{eq:s_tt_gen} \end{equation}

            \begin{equation} \sigma_{\phi \phi} = \sigma_{0}\left\{\text{sin}^2 \phi + \frac{1}{2\left(7-5\nu\right)}\frac{R^3}{r^3}\left[ 9 - 15\nu - 12\frac{R^2}{r^2} - \left(5 - 10\nu - 21\frac{R^2}{r^2}\right)\text{sin}^2\phi\right]\right\} = \sigma_{0} K_{\phi \phi}\left(\phi,r\right) \label{eq:s_pp_gen} \end{equation}

            \begin{equation} \sigma_{r \phi} =  \frac{ -\sigma_0}{2}\text{sin}2\phi \left\{ 1 + \frac{1}{2 (7-5\nu)} \frac{R^3}{r^3} \left[ 10\left(1+\nu\right) -24 \frac{R^2}{r^2} \right] \right\} = \sigma_{0} K_{r \phi}\left(\phi,r\right) \label{eq:s_rp_gen} \end{equation}

            This solution can then be superposed to determine the solution under uniaxial strain conditions which are characteristic for plate impact experiments. The longitudinal stress $\left(\sigma_{L}\right)$, is taken as the magnitude of the shock stress in the corresponding experiment, and the transverse, confining stress $\left(\sigma_{T}\right)$ is imposed in the lateral directions, as depicted in \cref{fig:ElasticDiagram}\hyperref[fig:ElasticDiagram]{c}. The confining stress, $\sigma_{T}$, is calculated based on the elastic Poisson's ratio, $\nu$, \begin{equation} \sigma_{T} = \frac{\nu}{1-\nu}\sigma_{L} \label{eq:LatStress}. \end{equation}
            The elastic confining stress also provides a good estimate for the lateral confining stress which is present in the shock experiments, the effect of which is discussed in \cref{sec:Results-Deformation}. The superposition procedure using the fundamental solution \cite{Southwell1926Concentration} in \cref{eq:s_rr_gen,eq:s_tt_gen,eq:s_pp_gen,eq:s_rp_gen} leads to the solution in \cref{eq:s_rr_sup,eq:s_tt_sup,eq:s_pp_sup,eq:s_rp_sup,eq:s_rt_sup} for the uniaxial strain loading of the plate impact experiments ($\sigma_{11}=\sigma_L$, $\sigma_{22}=\sigma_{33}=\sigma_T$). 
            \begin{equation} \sigma_{rr}(\sigma_{L},\phi,r) = \sigma_{L} K_{rr}\left(\phi,r\right) + \sigma_{T} K_{rr}\left(\phi - \frac{\pi}{2},r\right) + \sigma_{T} K_{rr}\left(\phi = \frac{\pi}{2},r\right) \label{eq:s_rr_sup} \end{equation} 
    
            \begin{equation} \sigma_{\theta \theta}(\sigma_{L},\phi,r) = \sigma_{L} K_{\theta \theta}\left(\phi,r\right) + \sigma_{T} K_{\theta \theta}\left(\phi - \frac{\pi}{2},r\right) + \sigma_{T} K_{\phi \phi}\left(\phi = \frac{\pi}{2},r\right) \label{eq:s_tt_sup} \end{equation} 
            
            \begin{equation} \sigma_{\phi \phi}(\sigma_{L},\phi,r) = \sigma_{L} K_{\phi \phi}\left(\phi,r\right) + \sigma_{T} K_{\phi \phi}\left(\phi - \frac{\pi}{2},r\right) + \sigma_{T} K_{\theta \theta}\left(\phi = \frac{\pi}{2},r\right) \label{eq:s_pp_sup} \end{equation}
            
            \begin{equation} \sigma_{r \phi}(\sigma_{L},\phi,r) = \sigma_{L} K_{r \phi}\left(\phi,r\right) + \sigma_{T} K_{r \phi}\left(\phi - \frac{\pi}{2},r\right) \label{eq:s_rp_sup} \end{equation}
    
            \begin{equation} \sigma_{r \theta}(\sigma_{L},\phi,r) = \sigma_{T} K_{r \phi}\left(\phi = \frac{\pi}{2},r\right) \label{eq:s_rt_sup} \end{equation}

            Finally, the stress state is converted to Cartesian coordinates and rotated to the principal frame ($\sigma_{\text{I}}$, $\sigma_{\text{II}}$, $\sigma_{\text{III}}$), from which Tresca shear stress and the direction of maximum shear can be determined. The stresses are also converted to strains through the generalized Hooke's law, 
            \begin{equation} \varepsilon_{ij} = \frac{1}{E}\left[\left(1+\nu\right)\sigma_{ij} - \nu\delta_{ij}\sigma_{kk}\right]. \label{eq:Hookes} \end{equation}
            The results derived in this section are used in \cref{fig:LineSlice,fig:SpacingPathsASB,fig:FracturePathArrest} and the corresponding discussion and analysis.

        \end{widetext}

\section{Effect of DIC subset size on measured strain in shear bands} \label{sec:AppendixDIC}
        
\begin{figure*}[htpb]
            \centering
            \includegraphics[width=1.0\textwidth]{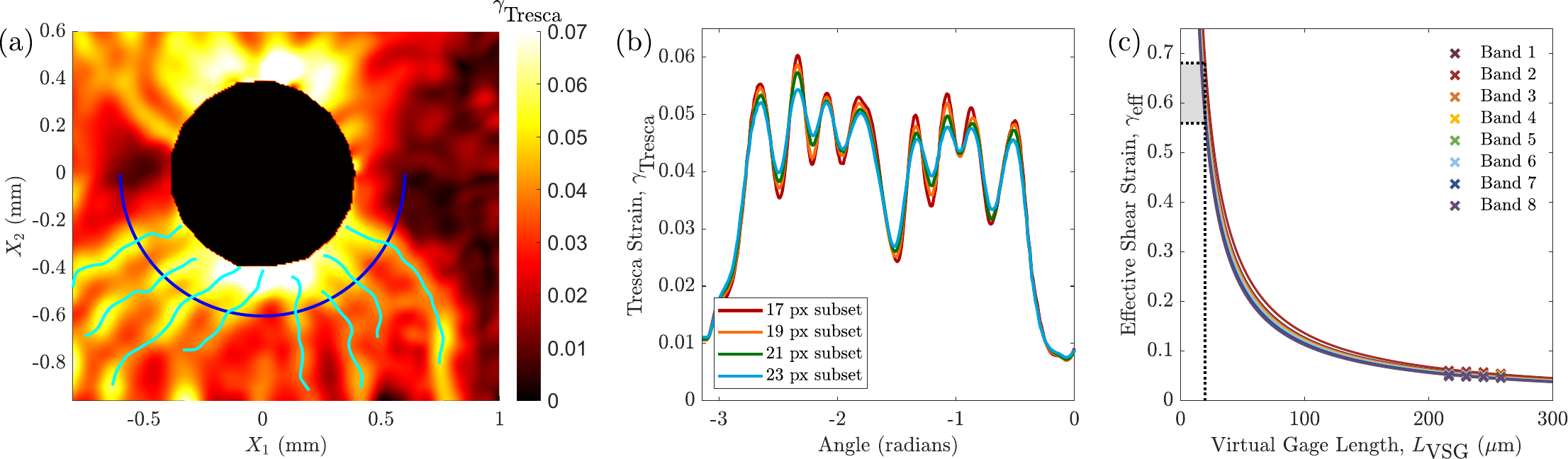}
            \caption{DIC subset size analysis for Pore--0.6. (a) Tresca shear strain contours at $t=0.7\,\mu$s are shown in the reference (undeformed) configuration. The semicircular blue curve represents the contour (at $r=0.6\,$mm) along which the full-field shear strain data is projected in 2D. (b) Projection of shear strain along semicircular contour for various subset sizes (strain window and step size are fixed at 15 and 1 pixels respectively). Lack of convergence for small subset size indicates physical nature of features and determines the actual strain to be larger than calculated. (c) Extrapolation of DIC measurements to estimate actual shear strain magnitude. }
            \label{fig:SubsetSizeAnalysis}
        \end{figure*}

        Digital image correlation (DIC) has an inherent filtering characteristic in the method, which must be taken into account, especially when investigating features of very fine spatial scale, as is the case when dealing with shear localization. To properly investigate the role of filtering on the DIC measurements and to make a comparison to the actual physical strain in the deformed body, one must carry out a DIC subset size analysis. The general idea is to compare the DIC measurement results when using various correlation settings, and compare the convergence of the results as a function of the filter size. The effective filter size can be summarized by the virtual strain gage length \cite{Reu2015VSG} (\cref{eq:VSG}), $$L_{\text{VSG}} = (SW-1)ST + SS.$$ If the calculated strain magnitude converges for a sufficiently small virtual strain gage length, then it is considered the actual strain magnitude. However, if no convergence is reached, then the largest calculated strain is taken as a lower bound for the actual strain magnitude, and the actual feature size is considered to be smaller than the smallest virtual strain gage length used. 
        In the case of the shear bands in Pore--0.6, the shear strain along a semicircular contour is considered, as shown in \cref{fig:SubsetSizeAnalysis}\hyperref[fig:SubsetSizeAnalysis]{a}. When projected into two dimensions in \cref{fig:SubsetSizeAnalysis}\hyperref[fig:SubsetSizeAnalysis]{b}, the shear bands can be seen as peaks, which makes visual inspection for various virtual strain gage lengths possible. It is observed that convergence is not achieved, indicating the bands are indeed a physical feature with large strain magnitude and small thickness, and also not a DIC artifact. Specifically, it is clear that the features are smaller than $217\,\mu$m and have strain magnitudes generally larger than the measured $6\%$ strain. Because DIC strain measurements scale as $\gamma \propto 1/L_{\text{VSG}}$, one can roughly estimate the actual strain magnitude via extrapolation, if the feature length scale is known. In this case, the size of the shear band is taken to be $20\,\mu$m based on prior postmortem measurements of plugged PMMA samples \cite{Rittel2008DynamicFlow}. Further assuming uniform strain in the shear band, one can fit a curve to the DIC measurements and determine the intersection at $L_{\text{VSG}}=20\,\mu$m, as is shown in \cref{fig:SubsetSizeAnalysis}\hyperref[fig:SubsetSizeAnalysis]{c}. Such a procedure suggests the actual strain magnitude ($\sim 60\%$) to be on the correct order of magnitude expected for an ASB. While this estimate of actual shear strain is admittedly a large extrapolation, it is not suggested as a quantitative measurement of strain in the shear bands. The purpose of this analysis is to demonstrate the importance of DIC correlation and filtering parameters, and to illustrate the particularly large influence of filtering on fine features such as adiabatic shear bands. The analysis shows that the DIC clearly captures a significantly filtered measurement of the deformation field, including distinct bands which are likely to have an actual strain on the same order as that of adiabatic shear bands.

\end{appendices}

\newpage
\newpage

\bibliography{References}

\end{document}